\documentclass[prd,aps,showpacs,preprintnumbers,amssymb,floats,
twocolumn,superscriptaddress,notitlepage,a4paper,
tightenlines
]{revtex4}
\usepackage{epsfig}
\usepackage{graphicx}% Include figure files
\usepackage{dcolumn}% Align table columns on decimal point
\usepackage{bm}% bold math

\def\ga{\mathrel{\raise.3ex\hbox{$>$\kern-.75em\lower1ex\hbox{$\sim$}}}}
\def\la{\mathrel{\raise.3ex\hbox{$<$\kern-.75em\lower1ex\hbox{$\sim$}}}}
\def\beq{\begin{equation}}
\def\eeq{\end{equation}}
\def\bea{\begin{eqnarray}}
\def\eea{\end{eqnarray}}

\begin{document}

\preprint{IPPP/02/51}
\preprint{DCPT/02/102}
\preprint{SISSA/61/2002/EP}

\title{Leptogenesis and rescattering in supersymmetric models}

\author{Lotfi Boubekeur} 
\affiliation{ SISSA-ISAS, Via Beirut 4, I-34013, and INFN, Sezione di Trieste, Trieste, Italy}
\author{Sacha Davidson}
\affiliation{ Dept. of Physics, University of Durham, DH1 3LE, United Kingdom}
\author{Marco Peloso}
\affiliation{ Physikalisches Institut, Universit\"at 
Bonn, Nussallee 12, D-53115 Bonn, Germany}
\author{Lorenzo Sorbo}
\affiliation{ GReCO, Institut d'Astrophysique de Paris, C.N.R.S., \\98bis Boulevard Arago, 75014 Paris, France}

\begin{abstract}
The observed baryon asymmetry of the Universe can be due to the $B-L$
violating decay of heavy right handed (s)neutrinos. The amount of the
asymmetry depends crucially on their number density. If the (s)neutrinos
are generated thermally, in supersymmetric models there is limited
parameter space leading to enough baryons. For this reason, several
alternative mechanisms have been proposed. We discuss the nonperturbative
production of sneutrino quanta by a direct coupling to the inflaton. This
production dominates over the corresponding creation of neutrinos, and it
can easily (i.e. even for a rather small inflaton-sneutrino coupling) lead
to a sufficient baryon asymmetry. We then study the amplification of MSSM
degrees of freedom, via their coupling to the sneutrinos, during the
rescattering phase which follows the nonperturbative production. This
process, which mainly influences the (MSSM) $D-$flat directions, is very
efficient as long as the sneutrino quanta are in the relativistic regime.
The rapid amplification of the light degrees of freedom may potentially
lead to a gravitino problem. We estimate the gravitino production by means
of a perturbative calculation, discussing the regime in which we expect it
to be reliable.
\end{abstract}

\maketitle

\section{Introduction and summary}The generation of the Baryon Asymmetry of the Universe (BAU)
\cite{Buchmuller:2000wq} represents one of the puzzles of Cosmology. Three
ingredients are required \cite{Sakharov:1967dj} to achieve this task:
baryon  number violation, $C$ and $CP$ violation, and departure from
thermal equilibrium. The baryon number violation can be challenging to
implement, because it must be consistent with the current lower bound on
the proton lifetime, $\tau_p \ga 10^{32}$ years~\cite{pdg}. The Standard
Model (SM) is a $C$ and $CP$ violating theory, and  contains
non-perturbative $B+L$ violating interactions (sphalerons)
\cite{sphalerons}---which are rapid in the early Universe but unable to
mediate proton decay. However, it seems difficult to use this baryon
number violation to create the asymmetry in the SM \cite{Kajantie:1995kf}
and its more popular extensions \cite{Carena:1997ki}. An attractive
alternative is to generate a lepton asymmetry \cite{FY} in some $C$, $CP$
and lepton number ($L$)-violating out-of-equilibrium interaction, and then
allow the sphalerons to reprocess part of it into a baryon asymmetry. An
appealing feature of this scenario is that while neutrino masses are 
experimentally observed \cite{SK} (and are $L$-violating, if they are
Majorana), there is still no evidence for baryon number violation.

The above idea is naturally implemented \cite{FY} in the context of the
see-saw \cite{seesaw}, which  is a minimal mechanism for generating
neutrino masses much smaller than the ones of the charged leptons. Three
right-handed (r.h.) neutrinos $N_i$ are added to the SM particle content,
given Yukawa interactions with the lepton and Higgs doublets, and large
Majorana masses. This gives the three light neutrinos very small masses,
due to their small mixing with the heavy r.h. neutrinos through the Dirac
mass. Grand Unified Theories (GUT) and their supersymmetric versions, that
constitute natural candidates for the Physics beyond the SM, often
contain  r.h. neutrinos in their particle content. In this paper we
consider the supersymmetric version of the see-saw mechanism, which is
theoretically attractive because it addresses the hierarchy between the
Higgs and r.h. neutrino masses.

The r.h. (s)neutrinos of the see-saw can generate the BAU via leptogenesis,
in a three steps process \cite{FY,boltzmann,both,review}. First, some
($CP$ symmetric) number density of (s)neutrinos is created in the early 
Universe. Then, a lepton asymmetry is generated in their $CP$ violating
out-of-equilibrium decay. Finally, the lepton asymmetry is partially
reprocessed into a baryon one by the $B+L$ violating interactions,
provided it is not washed out by lepton number violating scatterings. In
this paper, we are mostly interested in the first step, although in the
next section we will also briefly review the decay and washout processes.

The most straightforward and cosmological model-independent mechanism  to
generate r.h. (s)neutrinos is via scattering in the thermal
bath~\cite{review}. However, as discussed in section~\ref{sectiontwo}, the
parameter space available is restricted in supergravity-motivated models.
Indeed, unless some enhancement of the $CP$ asymmetry characterizing the 
r.h. (s)neutrino decay is present (which occurs for example if they are
nearly degenerate in mass) the generation of a sufficient lepton number 
poses a rather strong lower bound on their mass~\cite{HMY,FHY,sacha}
(see also \cite{MP}). 
For thermal
production, this translates into a lower bound on the reheating
temperature $T_{RH}$ of the thermal bath. On the other hand, if
supergravity is assumed, $T_{RH}$ cannot be taken arbitrarily large
without leading to an overproduction of gravitinos
\cite{wei,grinfla,grreh,grnuc,boundtr,grath}. The two requirements are
compatible only provided a nearly maximal $CP$ asymmetry (again, banning
any possible enhancement from mass degeneracy) is present in the r.h.
(s)neutrino decay. If this is not the case, alternative mechanisms of
production for the r.h. (s)neutrino have to be considered.

As remarked in~\cite{muya}, leptogenesis can be achieved if at least one
r.h. sneutrino has a smaller mass than the Hubble parameter, {\it i.e.}
\footnote{We will use $N$ as a shorthand for the superfield, its scalar
and fermionic component. We explicitly refer to the particle type when
this could cause a confusion.} $M_N<H$, during inflation. In this case,
quantum fluctuations of this sneutrino component are produced during the
inflationary expansion, and amplified to generate a classical condensate.
The decay of the condensate eventually generates the required lepton
asymmetry. The above requirement $M_N<H$ is not trivially satisfied in a
supergravity context, since supergravity corrections typically provide a
mass precisely of order $H$ to any scalar field of the model \cite{DRT}.
In this case, however, a suitable choice of the K\"ahler potential can
induce a negative mass term $m^2_{\rm ind} \simeq -H^2$, so that a large
expectation value will be generated for the sneutrino component during
inflation~\cite{DRT}. This also leads to the formation of a condensate
during inflation, and to successful leptogenesis as in the previous case.

Large variances can be produced during inflation if the sneutrinos are not
too strongly coupled to the inflaton field $\phi$, since this would
generate a high effective mass which could fix $\langle {\tilde N} \rangle
= 0$ during inflation. However, if one of the r.h. (s)neutrinos is coupled
to the inflaton, there is the obvious possibility that a sufficient amount
of (s)neutrino quanta is generated when the inflaton decays. Quite
remarkably, for a rather wide range of models this decay occurs in a
nonperturbative way~\cite{nonthana1.1,preh} (this is known in the
literature as {\it preheating}~\cite{preh}). In models of chaotic
inflation~\cite{chin}, this is due to the coherent oscillations of the
inflaton field, which can be responsible for a parametric amplification of
the bosonic fields to which the inflaton is coupled~\footnote{A
nonperturbative inflaton decay also occurs for hybrid
inflation~\cite{tac,abc,cpr}. However, we will not discuss this
possibility here.}. It is important to remark that this resonant
amplification does not require very high couplings between the inflaton
and the produced fields. For a coupling of the form $(g^2/2) \, \phi^2 \,
N^2$ in the scalar potential,  resonant amplification of the field $N$
already occurs for $g^2 \ga 10^{-8}\,$ \cite{cldeca}, if the mass of $N$
is negligible at the end of inflation, and if a massive inflaton is
considered. For a massless inflaton ($V \left( \phi \right)= \lambda
\,\phi^4/4$), an efficient resonance is present also for much smaller
values of $g\,$ (we will show this explicitly in section $3$), since in 
this case the resonance is not halted by the expansion of the 
Universe~\cite{nonthana2}.

If the produced particle is very massive ($M_N \ga m_\phi$), the
effectiveness of the resonance becomes a highly model dependent issue. A
potential of the form $V \left( \phi \right) + M_N^2 \, N^2 /2 + g^2 \,
\phi^2 \, N^2 / 2\,$, has been considered in the literature mainly to
discuss the production of heavy bosons needed for GUT baryogenesis
\cite{gutpreh}. Working in the Hartree approximation, it has been
found~\cite{cldeca} that a resonance is effective only provided the
coupling $g^2$ satisfies $g^2 \ga 10^{-7} \left( M_N / m_\phi \right)^4
\,$. Taking into account all the other backreaction effects, a stronger
lower bound on $g$ has to be expected~\cite{preh,cldeca}, since the latter
typically limits the growth of the fluctuations amplified by the
resonance.

Very different bounds can be expected for different potentials. Consider
for example $V \left( \phi \right) + \left( M_N + g \, \phi \right)^2 \,
N^2/2 \,$. In this case, due to the high initial amplitude of the inflaton
oscillations, the total mass of $N$ can vanish at some discrete points
even for a coupling as small as $g^2 \sim 10^{-10} \, \left( M_N / m_\phi
\right)^2 \,$. Whenever $M_N + g \, \phi = 0\,$, parametric amplification
of $N$ occurs. Thus, the lower bound valid for the previous potential is 
considerably weakened. Although this second choice of the potential may
seem {\it ad hoc}, we note that it is the one which arises in
supersymmetric models if both the r.h. sneutrino mass and interaction with
the inflaton are encoded in the superpotential, $W \left( N \right)
\supset M_N \, N^2 + g \, \phi \, N^2 \,$. We regard this as a very
natural possibility.

The idea of a nonperturbative production associated to the vanishing of
the total mass has been applied to leptogenesis in~\cite{nonthermal}. The
analysis of~\cite{nonthermal} focused on the production of r.h. neutrinos,
with a mass term of the form $\left( M_N + g \, \phi \right) {\bar N}
N\,$. From the results of~\cite{nonthermal}, and from the analytical
computations of~\cite{ps1}, it can be shown that a sufficient lepton
asymmetry is generated if the mass of the r.h. neutrinos is higher than
about $10^{14} \,$ GeV, and if their coupling to the inflaton satisfies $g
\ga 0.03\,$ (we will derive these bounds in section $2.3$). Here we note
that this high coupling can in principle destabilize through quantum
effects the required flatness of the inflaton potential. This, in addition
to the strong hierarchy between the r.h. neutrino mass and the electroweak
scale, motivates the study of the supersymmetrized version of the
mechanism proposed in~\cite{nonthermal}.

One of our aims is to show explicitly that, in the supersymmetrized
version of the above model, the nonperturbative production of the r.h.
sneutrinos is much more efficient than the one of the neutrinos. Due to
supersymmetry, the inflaton couples with the same strength both to the
r.h. neutrinos and to the sneutrinos, so that if the former are produced
at preheating this will also occur for the latter. However, while
production of fermions is limited by Pauli blocking, the production of
scalar particles at preheating is characterized by very large occupation
numbers. This high production has typically a big impact on the dynamics
of the inflaton field. The most immediate backreaction effect is the
generation of an effective potential for the zero mode of the inflaton.
This effective potential, taken into account in the Hartree
approximation~\cite{preh}, is typically comparable with or even dominant
over the tree level potential $V \left( \phi \right)\,$. There are however
two equally important effects which are beyond the Hartree approximation.
The first is due to the scatterings of the produced quanta against the
zero mode of the inflaton. This destroys the coherence of the
oscillations, thus ending the resonant production characterizing the early
stage of preheating~\cite{preh}. The second is the amplification of all
the other fields to which the produced quanta are coupled. This is a very
turbulent process, dominated by the nonlinear effects caused by the very
high occupation numbers of the fields involved. As a result, all these
mutually interacting fields are left with highly excited spectra far from
thermal equilibrium~\cite{felderkofman}. Both these effects are denoted as
{\it rescattering}~\cite{cldeca}.~\footnote{In some works, the term {\it
rescattering} refers only to the scatterings of the produced quanta
against the zero mode of the inflaton. Here we keep the original meaning
given in~\cite{cldeca}.}

Rescattering strongly affects some of the outcomes of the analytical
studies of preheating of bosons, which hardly go beyond the Hartree
approximation. For this reason, the results presented in our work are
obtained with numerical simulations on the lattice. More precisely, the
code ``LATTICEEASY''~\cite{latticeeasy}, by G.~Felder and I.~Tkachev, has
been used (details are given in section $3$). Full numerical calculations
on the lattice are however rather extensive. We have found that the
necessary computing time is reduced in the conformal case, that is with
the inflaton potential $V \left( \phi \right) = \lambda \, \phi^4 /4\,$,
and with a r.h. sneutrino mass which is negligible during the early stages
of preheating. For this reason, in our computations we fixed $M_N =
10^{11}\,$ GeV, which is smaller than the Hubble parameter during
inflation, but still high enough to require a nonthermal production of the
sneutrinos. The numerical results show a very efficient production of r.h.
sneutrinos and inflaton quanta at preheating/rescattering. Even for a
coupling inflaton-sneutrino as small as $g^2 \sim  {\rm few} \times
10^{-12}\,$, the produced quanta come to dominate the energy density of
the Universe already within about the first $5$ e-folds after the end of
inflation. In particular, the energy density stored in sneutrinos is
typically found to be a fraction of order one of the total energy density,
so that a sufficient leptogenesis is easily achieved at their decay.

R.h. sneutrinos are coupled to Higgs fields and left handed (l.h.) leptons
through the superpotential term $h \, N \, H \, L \subset W$ (responsible
for the Yukawa interaction which provides a Dirac mass to the neutrinos).
Thus, one may expect that quanta of the latter fields are amplified by the
rescattering of the r.h. sneutrinos produced at preheating. We study this
possibility in section $4\,$, showing that indeed the amplification occurs
for a wide range of values of the coupling $h\,$. Part of the analysis
follows the detailed discussion on rescattering given
in~\cite{felderkofman}, where the numerical code~\cite{latticeeasy} used
here was also employed. However, the analysis of~\cite{felderkofman} is
focused on the production of massless particles, while we show that the
non vanishing mass of the sneutrinos can have some interesting
consequences. More precisely, when the sneutrino quanta become non
relativistic (let us denote by ${\hat \eta}$ the time at which this
happens) their rescattering effects become much less efficient. Thus, a
strong amplification of the MSSM fields at rescattering can take place
only if the coupling $h$ is sufficiently large so that the amplification
occurs before ${\hat \eta}\,$. As a consequence, for massive sneutrinos
and for small values of $h\,$, the number of MSSM quanta produced at
rescattering is an increasing function of $h\,$. However, the production
is actually {\it disfavored} when the coupling $h$ becomes too high. This
is simply due to energy conservation, since the energy associated to the
interaction term between the sneutrino and the MSSM fields cannot be
higher than the energy initially present in the sneutrino distribution
(equivalently, one can say that, for a too high coupling $h\,$, the non
vanishing value of the sneutrinos gives a too high effective mass to the
MSSM fields, which prevents them from being too strongly amplified).
Posing quantitative bounds on the coupling $h$ would require some better
(analytical) understanding of the details of rescattering than we
presently have. However, the numerical results shown in section $4$ may
give an idea of the expected orders of magnitude.

An important remark is in order. When we speak about the amplification of
MSSM fields coupled to the r.h. sneutrinos we have actually in mind
amplification of $D-$flat directions (let us generally denote them by
$X$). Indeed, $D-$terms provide a potential term of the form $\Delta V
\sim g_G^2 \vert Y \vert^4$ for any scalar non flat direction $Y\,$. Since
$g_G$ is a gauge coupling ($g_G = O \left( 10^{-1} \right)$), we expect
such terms to prevent a strong amplification of $Y\,$, again from energy
conservation arguments. Another important issue which emerges when gauge
interactions are considered is whether gauge fields themselves are
amplified at rescattering. We believe that, at least in the model we are
considering, also the amplification of gauge fields will be rather
suppressed. The scalar distributions amplified at rescattering break much
of the gauge symmetry of the model. This gives the corresponding gauge
fields an effective mass in their dispersion relation (analogous to the
thermal mass acquired by fields in a thermal bath) of the order $m^2 \sim
g_G^2 \langle X^2 \rangle\,$. As we extensively discuss in the paper, in
the class of models we are considering the nonthermal distributions formed
at rescattering are characterized by a typical momentum several orders of
magnitude smaller than this mass scale. For this reason, one can expect
that such heavy gauge fields cannot be strongly amplified.~\footnote{Gauge
fields which are not coupled to the fields generated at rescattering will
not acquire this high effective mass. However, being uncoupled, they will
not be amplified either.} In our opinion, an explicit check of these
conjectures by means of numerical simulations could be of great interest,
especially considering the great importance that gauge fields could have
for the thermalization of the scalar distributions.

To conclude, we discuss the production of gravitino quanta from the scalar
distributions generated at rescattering. We already mentioned that in
order to avoid a thermal overproduction of gravitinos an upper bound has to
be set on the reheating temperature $T_{RH}$ of the thermal bath,
$T_{RH}
 \la {\rm few} \times 10^{10} \,$ GeV~\cite{grath}. The
requirement of a low reheating temperature can be seen as the demand that
the inflaton decays sufficiently late, so that particles in the thermal
bath have sufficiently low number densities and energies when they form.
If $H \simeq 10^{12} \,$ GeV at the end of inflation, and if the scale
factor $a$ is normalized to one at this time, the generation of the
thermal bath cannot occur before $a \simeq 10^7\,$. Gravitino
overproduction is avoided by the fact that in the earlier times most of
the energy density of the Universe is still stored in the coherent
inflaton oscillations. On the contrary, we have already remarked that
preheating/rescattering lead to a quick depletion of the zero mode in the
first few e-folds after the end of inflation.~\footnote{The situation is
even more enhanced for hybrid inflationary model, in which the energy
density of the zero modes of the scalars gets dissipated within their
first oscillation~\cite{tac,abc,cpr}.}

The question whether also the distributions formed at rescattering may
lead to a gravitino problem is thus a very natural one, and section $5$ of
the paper is devoted to some considerations on this regard.~\footnote{We
acknowledge very useful discussions with Patrick B. Greene and Lev Kofman
on this issue.} To provide at least a partial answer to this question, we
distinguish the period during which rescattering is actually effective
from the successive longer thermalization era. The computation of the
amount of gravitinos produced during the earlier stages of rescattering
appears as a very difficult task. The numerical simulations valid in the
case of bosonic fields indicate that a perturbative computation (with
dominant $2 \rightarrow 2$ scatterings taken into account) can hardly
reproduce the numerical results, and that probably $N \rightarrow 2$
processes ($N > 2$) have also to be taken into account (we discuss this
point in more details in section $4$). It is expected that the same
problem will arise also for the computations of the quanta of gravitinos
produced by the scalar distributions which are being forming at this
stage. The end of rescattering/beginning of the thermalization period is
instead characterized by a much slower evolution of the scalar
distributions. In particular, the total occupation number of all the
scalar fields is (approximatively) conserved, which is
interpreted~\cite{felderkofman} by the fact that $2 \rightarrow 2$
processes are now determining the evolution of their distributions.
Motivated by this observation, we assume that $2 \rightarrow 2$
interactions are also the main source of production for gravitinos from
this stage on.
%\begin{widetext}

\begin{table*}
\renewcommand{\arraystretch}{1.45}
\begin{tabular}{|l|c|c|c|} \hline
Mechanism & $N$ Yukawa $h$ & $N$ mass & $\phi$-$N$ coupling \\ \hline
Thermal &  $10^{-5} {\rm eV}< \tilde{m}_1 <10^{-3} {\rm eV}$ & $10^9 {\rm GeV} \la M_1\la T_{RH}$&  irrelevant \\
\phantom{Thermal} & \phantom{ $10^{-5} {\rm eV}< \tilde{m}_1 <10^{-3} {\rm eV}$} & \phantom{ $10^8 {\rm GeV} < M_1<
T_{RH}$ }& \phantom{irrelevant} \\
Affleck--Dine & $10^{-9} {\rm eV} < m_{\nu_1}< 10^{-4} {\rm eV}$ & $M_i<H_{\rm {infl}}$ & $
\left\{
\renewcommand{\arraystretch}{1}
 \array{l}
    M_i^{\rm eff}<H_{\rm infl}\\
    \left(M_i^{\rm eff}\right)^2<0
 \endarray
\right.
\renewcommand{\arraystretch}{1.45}
$
 \\
\phantom{Thermal} & \phantom{ $10^{-5} eV< \tilde{m}_1 <10^{-3} eV$} & \phantom{ $10^8 {\rm GeV} < M_1<
T_{RH}$ }& \phantom{irrelevant} \\
$
\left.
\renewcommand{\arraystretch}{1}
 \array{l}
    {\mathrm {Pert.}} \phi \\
    {\mathrm{decay}}
 \endarray
\right\}
\renewcommand{\arraystretch}{1.45}
$
 &
$\Gamma_{LV} <  H(\tau_i)$
 &
$
 \renewcommand{\arraystretch}{1}
 \left\{
  \array{c}
    M_i < m_\phi/2 \\
    M_i > m_\phi/2
  \endarray
 \right\}
\renewcommand{\arraystretch}{1.45}
$
 &
$ \renewcommand{\arraystretch}{1}
  \array{c}
    BR(\phi \rightarrow N_i N_i) \sim 1 \\
    BR(\phi \rightarrow N_i^* N_i^*) \sim 1
  \endarray
\renewcommand{\arraystretch}{1.45}$\\
\phantom{Thermal} & \phantom{ $10^{-5} eV< \tilde{m}_1 <10^{-3} eV$} & \phantom{ $10^9 {\rm GeV} \la M_1\la
T_{RH}$ }& \phantom{irrelevant} \\
$
\renewcommand{\arraystretch}{1}
 \array{l}
    N\,{\mathrm {preheating}} \\
    {\mathrm {eq.~}(\ref{lag})}
 \endarray
\renewcommand{\arraystretch}{1.45}
$
& $\Gamma_{LV} < H(\tau_i)$   & $M_i \ga 10^{14}\,{\rm GeV}$ & $g_i \ga 0.03$ \\
\phantom{Thermal} & \phantom{ $10^{-5} eV< \tilde{m}_1 <10^{-3} eV$} & \phantom{ $10^8 {\rm GeV} < M_1<
T_{RH}$ }& \phantom{irrelevant} \\
$
\renewcommand{\arraystretch}{1}
 \array{l}
    \tilde{N}\,{\mathrm {preh./resc.}}\\
    {\mathrm {eq.~}(\ref{pot1})}
 \endarray
\renewcommand{\arraystretch}{1.45}
$
& $\Gamma_{LV} <  H(\tau_i)$ & $M_i\la g_i\, 10^{17}\,{\rm GeV}$ & $g_{i} \ga \sqrt{\lambda}$ \\ \hline
\end{tabular}
\caption{
%{\footnotesize
 Summary of parameters for which leptogenesis could 
work, for different r.h. (s)neutrino production mechanisms. In (s)neutrino
production mechanisms which do not require the Yukawa coupling
(non-thermal mechanisms), the constraint on the Yukawa matrix is that
lepton number violating interactions in the thermal soup be out of
equilibrium after the r.h. (s)neutrinos decay at $\tau_i$. This also
implies $M_i > T(\tau_i)$, and possibly additional constraints on L
violating processes mediated by $N_j, j \neq i$. Recall $\tilde{m}_i$
parametrises the $N_i$ decay rate, and is defined after eqn. (\ref{k}).
The Affleck--Dine mechanism proceeds through generation of large
expectation values either for a small~\cite{ad} or a tachyonic~\cite{DRT}
effective mass of sneutrinos $M^{\rm eff}$ during inflation. The
asymmetry  made by the perturbative decay of the inflaton can be generated
by the on-shell r.h. (s)neutrinos \cite{Ypert}, which subsequently decay,
or by the decay via off-shell r.h. (s)neutrinos ($N_i^*$) to Higgses and
leptons \cite{alma}. The properties of nonperturbative (s)neutrino
production analysed in the present paper is summarized in  the last two
lines ($\lambda$ is the inflaton self-coupling). Other scenarios for
nonperturbative production after the end of inflation can be envisaged,
with model--dependent results.}
\label{tab1}
\end{table*}
\renewcommand{\arraystretch}{1}
%\end{widetext}
In the thermal case, the gravitino production is dominated by processes
having a gravitationally suppressed vertex (from which the gravitino is
emitted) and a second vertex characterized by a gauge interaction with one
outgoing gaugino. However, we believe that in the present context these
interactions will be kinematically forbidden, due to the high effective
mass-squared that gauginos acquire from their interaction with the scalar
distributions (the argument follows the one already given for gauge
fields). Once again we notice that the system is still effectively
behaving as a condensate: the number densities of the scalar distributions
are set by the quantity $\sqrt{\langle X^2 \rangle}\,$, which is much
higher than the typical momenta of the distributions themselves. This
generates a high effective mass for all the particles ``strongly'' coupled
to these scalar fields. A further comparison with the case of a thermal
distribution may be useful: in the latter case both the typical momenta
and the effective masses are set by the only energy scale present, namely
the temperature of the system. As should be clear from the above
discussion (see also~\cite{felderkofman}), the thermalization of the
distributions produced at rescattering necessarily proceeds through
particle fusion. Only after a sufficiently prolonged stage of
thermalization, will the system be sufficiently close to thermodynamical
equilibrium so to render processes as the one discussed above
kinematically allowed.

In section $5$ we show that if this class of processes is indeed
kinematically suppressed, the production of gravitinos from the
distributions formed at rescattering is sufficiently small. However, we
remark that this analysis still leaves out the gravitino production which
may have occurred at the earlier stages of the rescattering period.
Whether this production may be sufficiently strong to overcome the limits
from nucleosynthesis remains an open problem.

Let us finally summarize the plan of the paper. In section $2.1$ we
introduce our notation and briefly discuss some neutrino low energy
phenomenology. In section $2.2$ we discuss leptogenesis with a thermal
production of the r.h. (s)neutrinos. Leptogenesis with a nonthermal
production of r.h. neutrinos is reviewed in the following subsection. The
supersymmetric version of this model is presented in section $3\,$,  where
we study the nonthermal production of sneutrino quanta. Section $4$  is
devoted to the amplification of the MSSM $D-$flat directions due to the 
rescattering of the r.h. sneutrino quanta. The discussion on the
gravitino  production is presented in section $5\,$, apart from a few
technical details which can be found in the appendix.

\section{See-Saw Phenomenology and Leptogenesis}
\label{sectiontwo}

In this section, we introduce our notation for the SUSY see-saw and
outline  its low-energy implications. The aim is to make contact between
realistic  see-saw models, and the one generation toy models in which we
will study  the sneutrino production. We discuss the lepton asymmetry that
can be produced in (s)neutrino decay, which implies a lower bound on the
mass of the lightest r.h. (s)neutrino. Then, we briefly review different
mechanisms for r.h. neutrino production, namely thermal and non thermal.
The terms neutrino and sneutrino will be used interchangingly in
discussing thermal production, which is similar for bosons and fermions.
Concerning the nonthermal case, instead, different results are obtained
for the two species, and in  section \ref{GPRT} we review the ones for the
neutrinos. Nonthermal production of sneutrinos is instead discussed in the
next section.

\subsection{General Framework}

\vskip 10pt

Let us consider the Minimal Supersymmetric Standard Model (MSSM) extended
with three r.h. neutrino superfields $N_i$ (sometimes called the minimal
supersymmetric see-saw model). The relevant couplings of the r.h.
neutrinos are given by the superpotential~\footnote{The superfield $N$
written in eqn. (1) actually denotes an anti-(s)neutrino. However, for
brevity we will refer to it as a (s)neutrino.}
\begin{equation}
\label{superpotential}
W_N=h_{ji} \,L_i\cdot H_u \, N_j+\frac{1}{2}\,M_{k}\,N^2_k\,,
\end{equation}
where $L_i$ and $H$ are the lepton and the Higgs doublets, respectively,
and $h$ is a $3 \times 3$ complex Yukawa matrix. We will neglect the
phases in our analysis of $N$ production, because $CP$ violation is not
required for this process. We work in the r.h. neutrino mass basis, where
the mass matrix $M$ is diagonal, and we disregard the possibility of
nearly degenerate r.h. (s)neutrinos~\cite{dege,CP,both} (i.e. we assume
that  the
difference of neutrino masses is of order their mass).
%not fine tuned to be comparable with
%their width).

The lepton asymmetry  produced in the decay of $N_i$ can be written
\begin{equation}
Y_L\equiv{N_L-N_{\bar{L}}\over s}=\epsilon_i\;{N_{N_i}\over s} \kappa_i,
\end{equation}
where $N_{N_i}$ is the total number density of the $i$th heavy (s)neutrino
species prior to its decay, $s$ is the entropy density at decay
\footnote{Any subsequent entropy production leads to further dilution of
the asymmetry.}, $\kappa_i$ parametrises washout effects due to subsequent
lepton number violating interactions, and  $\epsilon_i\,$ arises from the
$CP$ violation of the $N_i$ decay. It is given by~\cite{CP}
\begin{eqnarray}
\epsilon_i&\equiv\ &\frac{\Gamma\left(N_i\rightarrow L\,H\right)-
\Gamma\left(N_i\rightarrow
\bar{L}\,\bar{H}\right)}{\Gamma\left(N_i\rightarrow
L\,H\right)+\Gamma\left(N_i\rightarrow
\bar{L}\,\bar{H}\right)}\nonumber\\
&=&
\frac{1}{8\pi \left(hh^\dagger
\right)_{ii}}\sum_{j}{\rm {Im}} \left[
\left(hh^\dagger\right)^2_{ij}\right] f\left(\frac{M^2_j}{M^2_i}\right)
\label{eps1}
\end{eqnarray}
where
$f(x) = \sqrt{x} [ 2/(x-1) + \ln (1/x+1) ]$
 for hierarchical r.h. neutrino
masses.

We suppose for the moment that some number density of $N_i$ is produced in
the early Universe, and concentrate on how large an asymmetry can be
generated. The asymmetry $\epsilon_i$ is determined by the masses and
couplings of the r.h. (s)neutrinos, which are given in eqn.
(\ref{superpotential}). However, it can be related to, and therefore
constrained by, low energy observables.

The $CP$ asymmetry produced in the decay of a r.h. (s)neutrino can
conveniently be parameterized as
\begin{eqnarray}
\epsilon_i &=& \frac{3}{8\,\pi}\,\frac{M_i \,m_3}{\langle
H\rangle^2}\delta_{\rm CP} \nonumber\\ &\simeq&
10^{-6}\left(\frac{M_i}{10^{10}{\rm
{GeV}}}\right)\left(\frac{m_3}{0.05\,{\rm {eV}}}\right)\delta_{\rm CP}.
\label{param}
\end{eqnarray}
By using eqs.~(\ref{see-saw}) and (\ref{eps1}), it is possible to
show~\cite{HMY,FHY,sacha} that  for the case $\epsilon_1$, $\delta_{\rm CP}$
satisfies the  upper bound \footnote{We will use the parametrisation
(\ref{param}) for all the $\epsilon_i$, $ i = 1..3$. It is possible that
$\delta_{CP} \leq  1$  for $\epsilon_2$ and $\epsilon_3$ (assuming no
cancellations in the formulae), although this has not been shown.}
\begin{equation}
\vert \delta_{\rm CP} \vert \leq 1\,\,.
\label{bound}
\end{equation}
By combining the two last expressions, one finds an upper bound on the
parameter $\epsilon_1$ which scales linearly with the r.h. (s)neutrino
mass $M_1\,$. We will shortly see that this implies a lower bound on $M_1$
for leptogenesis to be viable.
The mass $m_3$ in equation (\ref{param}) denotes the mass of the heaviest 
left-handed neutrino. The light neutrino mass matrix is obtained by
integrating out the heavy r.h. neutrinos to give the see-saw formula
\begin{equation}
\label{see-saw}
m_\nu= - h^T\,M^{-1}\,h\,\langle H^0_u\rangle^2 \,\,.
\end{equation}
We will assume that the light neutrino masses $m_i$ are hierarchical, so
$m_{3} \simeq \sqrt{\Delta m^2_{atm}}$ \cite{SK}.

If $h$ is written in the charged lepton mass eigenstate basis (neutrino
flavour basis), then $m_\nu$ is diagonalised by the MNS matrix $U$
\cite{Maki:1962mu}, which can be written $ U=V\cdot {\mathrm 
 {diag}}(e^{-i\phi/2},e^{-i\phi'/2},1)$, with
%
%\beq
%\label{MNS}
%V=\pmatrix{c_{13}c_{12} & c_{13}s_{12} & s_{13}e^{-i\delta}\cr
%-c_{23}s_{12}-s_{23}s_{13}c_{12}e^{i\delta} & c_{23}c_{12}-s_{23}s_{13}s_{12}e^{i\delta} & s_{23}c_{13}\cr
%s_{23}s_{12}-c_{23}s_{13}c_{12}e^{i\delta} & -s_{23}c_{12}-c_{23}s_{13}s_{12}e^{i\delta} &
%c_{23}c_{13}\cr}.
%\eeq
%\beq
%$$
\begin{widetext}
\[
%\beq
%\label{MNS}
V=\left[ 
\begin{array}{ccc}
 c_{13}c_{12} & c_{13}s_{12} & s_{13}e^{-i\delta} \\
-c_{23}s_{12}-s_{23}s_{13}c_{12}e^{i\delta} 
  & c_{23}c_{12}-s_{23}s_{13}s_{12}e^{i\delta} & s_{23}c_{13} \\
s_{23}s_{12}-c_{23}s_{13}c_{12}e^{i\delta} & -s_{23}c_{12}-c_{23}s_{13}s_{12}e^{i\delta} &
c_{23}c_{13}  
\label{MNS}
\end{array} \right].
%$$
%\eeq
\]
%\eeq
\end{widetext}
In this matrix, $c_{12} = \cos \theta_{12}$, and so on. Atmospheric
\cite{SK,atmex} and solar \cite{SNO,solex} data imply that $\theta_{12}$
and $\theta_{23}$ are large, approaching $\pi/4$. $\theta_{13}$ is
constrained to be $\la 0.1$ by the CHOOZ experiment \cite{CH}.
In a supersymmetric scenario, there is additional information about $h$
and $M$ available in the slepton mass matrix. The neutrino Yukawa
$h^\dagger h$ appear in the renormalization group equations for the soft
slepton masses, and thereby induce flavour violating slepton mass terms
\cite{Borzumati:1986qx}: $[\tilde{m}_L^2]_{ij}$. 
In a simple-minded leading log approximation, these off-diagonal mass matrix elements are
\begin{equation}
[\tilde{m}_L^2]_{ij}   \simeq  \frac{(3m_0^2 + A_0^2)}{8 \pi} [V_L]_{ki}^*
[V_L]_{kj} h_k^2 \log \left( \frac{M_k}{M_{GUT}} \right)
\label{**}
\end{equation}
where $h_i$ are the eigenvalues of $h$, $m_0^2$ and $A_0$ are soft
parameters at the GUT scale, and we introduce a new matrix $V_L$ which
diagonalises $h^\dagger h$ in the charged lepton mass eigenstate basis
($V_L h^\dagger h V_L^\dagger = $ diagonal). The branching ratio for 
$\ell_j \rightarrow \ell_i \gamma$ can be roughly estimated as 
\cite{Borzumati:1986qx}:
\beq
{\rm{BR}}(\ell_j \rightarrow \ell_i \gamma) \propto
\frac{ \alpha^3}{G_F^2} \frac{|
[\tilde{m}_L^2]_{ij}|^2}{\tilde{m}_L^8} \tan^2 \beta
\eeq
where $\tilde{m}_L^2$ is the slepton mass scale. The experimental bound
$BR(\mu \rightarrow e \gamma) < 1.2 \times 10^{-11}$ \cite{mega} implies
$[\tilde{m}_L^2]_{\mu e} \la 10^{-3} \tilde{m}_L^2$, for $\tilde{m}_L
\simeq 100$ GeV. This constrains the angles in $V_L$ for given neutrino
Yukawas $h_i$.
It can be shown that  $M$ and $h$  have the same number of parameters  as
the weak scale neutrino and slepton mass matrices. Furthermore, in a SUSY
scenario with universal soft masses at the GUT scale, $M$ and $h$ can be
parametrised with $\tilde{m}^2_L$ and $m_\nu$ \cite{Davidson:2001zk}. The
r.h. neutrino masses and Yukawa couplings can  be therefore  reconstructed
({\it in principle}, but not in practise \cite{Davidson:2001zk}) from the
weak scale neutrino and sneutrino mass matrices, so that $\epsilon_i$ can
be expressed in terms of weak scale variables.
An analytic approximation for $\epsilon_1$ can be found in \cite{Davidson:2002em,Branco:2002kt}:
\beq
\epsilon_1 \simeq \frac{3 h_1^2}{8 \pi \sum_j |W_{1j}|^2  m_{j}^2}
 {\rm Im} \left\{ \frac{ \sum_k W_{1k}^{2} m_{k}^3 }
{ \sum_n W_{1n}^{2} m_{n} } \right\} ~~,
\label{epsapprox}
\eeq
where $m_i$ are the light neutrino
masses, $h_1$ is the smallest eigenvalue of  $h$,
and   $W = V_L U$
is the rotation from the basis where the $\nu_L$ masses are
diagonal to the basis
where ${ h^{\dagger}} {h}$ is diagonal.
$h_1$ is in practise unmeasurable;
however, if $h$ has a hierarchy similar to the up
Yukawa matrix $h_u$, then $h_1^2 \sim 10^{-8}$,
and $\epsilon_1$  will only be large enough if there is
some enhancement from the imaginary part.
There are two simple limits for the matrix $W$, which are
motivated by model building. The
first is $V_L \simeq 1$, and corresponds to an almost diagonal
slepton mass matrix (in the charged lepton mass eigenstate
basis). This means that the
large mixing observed in the MNS matrix $U$ must come
from the r.h. sector \cite{Smirnov:af}. The second option
is $ W \simeq 1$, so $V_L \simeq U^\dagger$. This would
arise if the large $\nu_L$ mixing is induced in the l.h. sector
\cite{Altarelli:1998ns}.
In the $V_L = 1$ case,  eqn. (\ref{epsapprox}) gives \cite{Davidson:2002em}:
\begin{widetext}
\bea
\label{eps_analytic1}
\epsilon_1 & \simeq & -\frac{3 h_1^2}{8 \pi D} ~
{\rm Im} \left\{\frac{
m_{1}^3 ~c_{13}^2 c_{12}^2 ~e^{ i \phi}+
m_{2}^3 ~c_{13}^2 s_{12}^2~ e^{ i \phi'}+
m_{3}^3 ~s_{13}^2 ~ e^{2 i \delta}}
{m_{1}~ c_{13}^2 c_{12}^2 ~e^{ i \phi}+
m_{2} ~c_{13}^2 s_{12}^2~ e^{ i \phi'}+
m_{3} ~s_{13}^2 ~ e^{2 i \delta}} \right\}  \\ \label{eps_analytic2}
& \simeq & -\frac{3 h_1^2}{4 \pi } \left\{
\left(\frac{m_{3}}{m_{2}}\right)^3 2 s_{13}^2 \sin(2 \delta -  \phi')-
 \frac{m_{1}}{m_{2}} \sin (\phi-\phi')
 \right\}  ~~,
\eea
\end{widetext}
where $D= m_{1}^2 ~c_{13}^2 c_{12}^2+ m_{2}^2~ c_{13}^2 s_{12}^2+ m_{3}^2
~s_{13}^2$, and in the second equation, the solar and atmospheric angles
have been taken to be $\pi/4$.  If we estimate the phases to be $O(1)$,
$h_1 \sim $ the up Yukawa, and $m_3^2/m_2^2 \sim \Delta m^2_{atm}/ \Delta
m^2_{sol}$, this gives $\epsilon \la 10^{-7} (s_{13}/.1)^2$, where we have
scaled the unmeasured angle $\theta_{13}$ by its upper bound.
\footnote{This estimate is not significantly changed if the angles in
$V_L$ are small compared to $ \theta_{13}$. % which %is constrained by the
CHOOZ experiment:   $ \theta_{13} \la .1$. If  $ \theta_{13} \ll .1$, but
$ \theta_{13} < [V_L]_{12}, [V_L]_{13} <  .1$, then the  formula for
$\epsilon$ is similar to (\ref{eps_analytic2}), with the replacement
$\theta_{13} \rightarrow \theta_{L1j}$ and $\delta \rightarrow
\varphi_{1j}$  (where $ [V_L]_{12} =  \cos \theta_{L13}\sin
\theta_{L12}e^{i \varphi_{12}} $, $ [V_L]_{13} =  \sin \theta_{L13}e^{i
\varphi_{13}}$).} This is barely large enough for thermal leptogenesis.
However, we remind that $h_1$ is unknown and it can well be $h_1 >
10^{-4}\,$.
The second case, where $W \simeq 1$, can arise if $M$ and
$h^\dagger h$ are almost simultaneously diagonalisable \footnote{The
``almost'' is important; if $ W = 1$, there is no $CP$
violation, so $ \epsilon = 0$.}.  For small angles in $W$,
the approximation for $\epsilon$ can be extracted from
(\ref{eps_analytic1}), replacing the angles of the MNS matrix
by the angles of $W$, and setting the cosines $\rightarrow 1$.
When $W \sim 1$, then $V_L \simeq U^\dagger$, so it is
the MNS angles that appear in equation (\ref{**}), and
$BR( \mu \rightarrow e \gamma) < 1.2 \times 10^{-11}$  \cite{mega} 
implies an upper bound on the CHOOZ angle $\theta_{13} < .02$
(for $\tilde{m}_{L} = 100$ GeV, $h_3 = 1$)\cite{Sato:2000zh}.
To conclude, we briefly comment on the parameter $\kappa_i$. After the
asymmetry is generated in the out-of-equilibrium decay of the r.h.
neutrino, lepton number violating interactions which could wash out the
asymmetry must  be out of equilibrium. This is a fairly straightforward
requirement when considering the decay of the lightest r.h. neutrino $N_1$
\cite{review}; it is more complicated in the case of $N_{2,3}$ decaying at
$T \ga M_1$ \cite{Barbieri:1999ma}. The fraction of the asymmetry which
survives these interactions is $\kappa_i \leq 1$.
\subsection{Thermal $N_i$ production}
We now consider the case where the lightest r.h. (s)neutrino $N_1$ is
thermally produced after $T_{RH}$ \footnote{We assume in this work
that $T_{RH}$ is ``large''; for a discussion of baryogenesis
in low-$T_{RH}$ models, see $e.g.$ \cite{toni}.}.
 With hierarchical r.h. neutrino masses,
one can typically assume the lepton asymmetry to be produced by the decay
of the lightest r.h. (s)neutrino $N_1\,$. As we shall see, this is a
self-consistent assumption, because $M_2$ and $M_3$ will turn out to be
larger than $T_{RH}$.
To generate a lepton asymmetry, the decay of the $N_1$  should proceed out
of equilibrium. More quantitatively, the ratio of the thermal average of
the $N_i$ decay rate and of the Hubble parameter at the temperature $T
\simeq M_i\,$,
\begin{equation}
\frac{\Gamma_{N_i}}{2\,H} \vert_{T=M_i} \equiv
\frac{\tilde{m}_i}{2 \times 10^{-3} \, {\rm {eV}}} \,\,,
\label{k}
\end{equation}
should be less  than unity
to have an unambiguously out-of-equilibrium decay.
The parameter $\tilde{m}_i$ is defined as
$\Gamma_i \langle H \rangle^2/M_i^2$, where
$\langle H \rangle$ is the Higgs vev.
However, $\tilde{m}_1$ cannot be taken too small if $N_1$ is
produced thermally~\cite{review}. Indeed the
quantity $\tilde{m}_1$ controls the
strength of the interactions of the $N_1$ with
MSSM degrees of freedom, and an efficient thermal production
via Yukawa interactions typically
requires $\tilde{m}_1 \ga \,10^{-5} $ eV.
To account for both these effects, $\Gamma_{N_1} \sim H(T = M_1)$ should 
be taken, so the decay is only barely out of equilibrium, and the final
lepton asymmetry has to be computed by integrating the full set of
relevant Boltzmann equations \cite{boltzmann,both}. These
computations show that a significant portion of the lepton asymmetry is
erased by lepton-number violating processes, and that only a fraction
$\kappa \la .1$ or less typically survives. Starting from $N_1$ in thermal
equilibrium at $T > M_1\,$, and collecting all the above informations,
the  final baryon asymmetry can be estimated to be
\begin{equation}
Y_B \simeq
10^{-10}\,\left(\frac{M_1}{10^{10}\,{\rm
{GeV}}}\right)\,\left(\frac{m_3}{0.05\,{\rm
{eV}}}\right)\,\left(\frac{\kappa}{0.1}\right)\,\delta_{\rm CP}\,\,.
\label{bth}
\end{equation}
This expression has to be compared with the baryon asymmetry required by 
Big Bang Nucleosynthesis $ Y_B\equiv N_B / s \sim (1.7 - 8)\times
10^{-11}$  \cite{keith,pdg}. As we have anticipated, we see that thermal
leptogenesis can be a viable mechanism if the mass of the r.h. neutrino
$N_1$ is sufficiently high. From eqn.~(\ref{bth}) we find the lower bound
$M_1 \ga \left( 10^9 - 10^{10} \right)\,$ GeV, although it is fair to say
that higher values are required if the bound~(\ref{bound}) is not
saturated. If $N_1$ is generated from the thermal bath, a reheating
temperature greater than $M_1$ is required. In the most favorable case,
the required value is only marginally compatible with the bound imposed by
gravitino overproduction from the thermal bath~\cite{grath}. To overcome
the potential conflict between the gravitino bound and the requirement of
a reheating temperature high enough for leptogenesis, the possibility of
producing right-handed neutrinos {\it non thermally} has been envisaged
\cite{nonthermal}. The following considerations will be focused on this
framework.
\subsection{Non thermal production of right handed neutrinos}
\label{GPRT}
Many alternatives to thermal (s)neutrino production have been considered
in the literature. We concentrate here on mechanisms that involve a direct
coupling of $N\,$ to the inflaton, although in the next section we will
comment on differences and similarities with the Affleck-Dine mechanism.
The strength of this interaction, relative to the inflaton coupling to
other degrees of freedom, is a free parameter; for appropriate values, a
lepton asymmetry of the correct magnitude can be produced.  The number
density of r.h. (s)neutrinos will also depend on the evolution of the
inflaton between the end of inflation and reheating. If the inflaton
decays perturbatively, right-handed neutrinos with masses less than half
the inflaton mass could be produced in the decay \cite{Ypert}. For heavier
r.h. neutrinos, one can also envisage the possibility that a sufficient
leptogenesis is generated in processes in which they mediate a
perturbative inflaton decay~\cite{alma}. In both cases, the final lepton
asymmetry will be proportional to the branching ratio of the inflaton into
(either on- or off-shell) neutrinos. A branching ratio of order one is
typically required.
Right-handed neutrinos with masses greater than that of the inflaton can
be produced at preheating, if their interaction with the inflaton is
strong enough. The production of heavy fermions (sneutrinos are discussed
in the next section) in an expanding Universe was first discussed in
ref.~\cite{nonthermal} (fermionic production in the conformal case was
first studied in~\cite{gk}), where a direct Yukawa coupling to the
inflaton $\phi$ was considered, and the simplest chaotic inflationary
scenario with a massive inflaton, $V \left( \phi \right) = m_\phi^2 \,
\phi^2 / 2 \;,\; m_\phi \simeq 10^{13} \, $ GeV, was assumed. The relevant
part of the lagrangian is
\begin{equation}
{\cal {L}}_{N,\,\phi}=\bar{N}\,\left(M+g\,\phi\right)\,N\,\,,
\label{lag}
\end{equation}
where $N$ is any one of the r.h. neutrinos. We assume that only one r.h.
neutrino generation plays an important role in the generation of a lepton
asymmetry, and therefore we drop the r.h. neutrino generation index for
the remainder of this section. The  generalization of the following
analysis to three generations is straightforward, at least as long as the
r.h. neutrino--inflaton coupling matrix $g$ is diagonal in the r.h.
neutrino mass basis (otherwise, the formalism of \cite{granonth2} should
be used).
After the end of inflation, the inflaton condensate $\phi$ oscillates
about the minimum of its potential with amplitude of a fraction of the
Planck mass $M_P \simeq 1.22 \cdot 10^{19} \, {\rm GeV} \,$. The total
effective mass of the fermion $M + g \, \phi \left( t \right) $ varies non
adiabatically in time, and this leads to a (non perturbative) production
of quanta of $N$. In particular, fermion production at preheating occurs
whenever the total effective mass crosses zero. As a consequence, fermions
with a mass up to
% 
%\begin{eqnarray}
\begin{equation}
\label{mmax}
M_{\rm{max}}\simeq 5\left(\frac{q}{10^{10}}\right)^{1/2}\times
10^{17}\,{\rm {GeV}}, \,
%\qquad 
%\nonumber
q\equiv
\frac{g^2\phi_0^2}{4m_\phi^2}\simeq 3g^2\,10^{10}\,\,
\end{equation}
%\end{eqnarray}
%
can be produced~\cite{nonthermal}, irrespective of the value of the
reheating temperature of the thermal bath which is formed at later times.
The abundance of neutrinos produced at preheating has been computed
analytically~\cite{ps1}, and it is most conveniently given in terms of the
ratio
%\begin{widetext}
\begin{equation}\label{ratio}
\frac{N_N}{\rho_\phi} \simeq \frac{1}{10^{10}\,{\rm{GeV}}}\,\frac{1.4\times
10^{-14}\,q}{M_{10}^{1/2}}\,\left[\log\left(1.7\times
10^3\,\frac{q^{1/2}}{M_{10}}\right)\right]^{3/2}\,\,,
\end{equation}
%\end{widetext}
%
where we have defined $M_{10}=M/\left(10^{10}\,{\rm{GeV}}\right)$. The
above formula is valid as long as the backreaction of the produced
neutrinos on inflaton dynamics is negligible, as it turns out to be the
case as long as $q\la 10^8$~\cite{nonthermal,ps1}. For larger values of
$q$, the effectiveness of preheating increases (by a factor up to about
$1.5$), and the above equation gives just a lower bound on $N_N$. In what
follows, we will conservatively assume $q<10^8\,$. \footnote{One should
also consider the perturbative decay of the inflaton quanta. Comparing
eqn.~(\ref{ratio}) with the number of r.h. neutrinos produced
perturbatively in one inflaton oscillation (i.e. the typical timescale for
preheating), one can however see that the perturbative production is
subdominant when kinematically allowed.}
For a massive inflaton, the ratio~(\ref{ratio}) is constant until the
inflaton condensate decays. If neutrinos decay before reheating has
completed, the resulting baryon asymmetry  reads
\begin{widetext}
\begin{eqnarray}\label{asinfdom}
Y_B &=& \frac{8}{23} Y_{B-L}=\frac{8}{23}\,\left(-\epsilon\,
N_N\,\frac{3}{4}\,\frac{T_{\rm RH}}{\rho_\phi} \right) \nonumber \\
&=&4 \cdot 10^{-14} \, \frac{q}{10^8} \, M_{10}^{1/2}\,
\frac{T_{\rm RH}}{10^9 \, {\rm GeV}} \, \frac{m_{\nu_3}}{0.05 \, {\rm eV}}
\, \delta_{CP} \, \, \left[ {\rm Log } \left( 1.66 \cdot 10^3
\, \frac{q^{1/2}}{M_{10}} \right) \right]^{3/2} \,\,.
\end{eqnarray}
\end{widetext}
The ratio of $N_N$ to the entropy is constant after reheating, unless it
decreased due to some subsequent entropy production \footnote{For instance, 
this is the case if the r.h. neutrinos lifetime is long enough for them to
come to dominate the energy density in the Universe, after reheating has
completed.}. For a given value of the parameter $q$, the baryon asymmetry
(18) is maximized when the mass M has its largest possible value (16). For
$T_{\rm {RH}}\simeq 10^9$~GeV, $m_{\nu_3}=0.05$~eV and $\delta_{CP}=1$,
imposing the condition (16) on eqn. (18) it is possible to see that the
generation of the observed baryon asymmetry requires $g$ larger than about
$0.03$. In particular, if we assume $q\simeq 10^8$, then the observed
baryon asymmetry can be obtained for a mass of the right--handed neutrino
of the order of $10^{14}-10^{15}$~GeV.
Production of r.h. neutrinos at preheating can generate a large enough
lepton asymmetry from neutrinos with $M_N > T_{RH}$, so constraints on
$T_{RH}$ do not translate into bounds on $M_N$. We notice however that the
large inflaton-neutrino coupling required ($g \ga 0.03$) can in principle
modify through quantum effects the small mass or self coupling parameters
characterizing the inflaton potential. This constitutes a further
motivation for considering supersymmetric models, as we do in the
remaining of this work. We will see that the production of the scalar 
partners of the r.h. neutrinos can significantly affect some of the above
considerations.
\section{Supersymmetric see-saw and nonthermal production of right handed sneutrinos}
As we have seen in the previous section, preheating can have very
important consequences for leptogenesis through the production of r.h.
neutrinos~\cite{nonthermal}. In supersymmetric extensions of the see-saw
model, the production of the supersymmetric partners of the neutrinos is
even more important. Due to supersymmetry, the inflaton couples with the
same strength both to the r.h. neutrinos and to the sneutrinos, so that if
the former are produced at preheating this will also occur for the latter.
However, while production of fermions is limited by Pauli blocking, the
production of scalar particles at preheating is characterized by very
large occupation numbers. As a consequence, the production of r.h.
sneutrinos can be expected to be more significant than the one of
neutrinos, as the numerical results presented below confirm.
Production of particles at preheating gives very model dependent results;
nevertheless, some general features can be outlined, and the whole process
can be roughly divided into three separate stages. The first of them is
characterized by a very quick amplification of the fields directly coupled
to the inflaton (and of the inflaton field itself, in the case of a
sufficiently strong self-interaction) to exponentially large occupation
numbers~\cite{preh}. Very rapidly, the system reaches a stage in which the
backreaction of the produced quanta, customarily denoted as
rescattering~\cite{cldeca}, plays a dominant role. In the case of
parametric resonance, the scatterings of the quanta against the zero mode
of the inflaton destroy the coherence of the oscillations, thus ending the
resonant production characterizing the early stage of
preheating~\cite{preh}. An equally important backreaction effect is the
amplification of all the other fields to which the produced quanta are
coupled. This is a very turbulent process, dominated by the nonlinear
effects caused by the very high occupation numbers of the fields involved.
As a result, all these mutually interacting fields are left with highly
excited spectra far from thermal equilibrium. The latter is actually
achieved on a much longer timescale, through an adiabatic (slow) evolution
of the spectra, which characterizes the third and final stage of the
reheating process.~\footnote{The thermalization of this system is a very
interesting issue,  which however we do not discuss in this work - see
\cite{felderkofman} for a more detailed study. Since in this case
thermalization proceeds via particle fusion, an important role may be
played by three or five point vertices, which shorten perturbative
estimates of the thermalisation timescale \cite{ur} (other recent
discussions on thermalization can be found in \cite{th1,th2,bec,adr}).}
The first stage of preheating is well understood. Particle production is
computed in a semi-classical approximation (for a rather general formalism
in the case of several coupled fields see~\cite{granonth2}), and
analytical solutions have been obtained in a broad class of
models~\cite{nonthana1.1,preh,nonthana1.2,nonthana2,gk,ps1}. Analytical
approximations break down when nonlinear processes become dominant.
However, the high occupation numbers of the scalar fields involved allow a
classical study of the system. Indeed, in the limit of high occupation
numbers quantum uncertainties become negligible, and quantum probabilities
show a classical (deterministic) evolution~\cite{qtocl}. The latter can be
better computed by means of lattice simulations in position
space~\cite{cldeca,latticeeasy,cpr}, where all the effects of backreaction
and rescattering are (automatically) taken into account. A detailed
discussion of rescattering and of the approach to thermal equilibrium has
been given in~\cite{felderkofman}, where the  code
``LATTICEEASY''~\cite{latticeeasy}, by G.~Felder and I.~Tkachev, has been
used. The numerical results presented in this paper are also obtained with
this code.
%\begin{widetext}
\begin{table*}
\renewcommand{\arraystretch}{1.45}
\begin{tabular}{|l|c|} \hline
 inflaton vev at the end of inflation  &$ \phi_0  \simeq M_P/3$ \\
inflaton self-coupling (eqn. (\ref{w1})) & $\lambda  \simeq  9 \times 10^{-14}$ \\
 neutrino-inflaton coupling (eqn. (\ref{w1})) &$ \tilde{g}  = g^2/\lambda $  \\
neutrino Yukawa coupling (eqn. (\ref{delw})) & $ \tilde{h}  = h^2/\lambda ~~
 ( \leftrightarrow \tilde{m} = \frac{h^2 \langle H_u \rangle^2}{M})$ \\
 X number density (eqn. (\ref{n})) & $N_{c,X} =$ [comov. num. den.]/[$\sqrt{\lambda} \phi_0]^3$ \\
 X ``mass'' (eqn. (\ref{mass})) & $m^2_{{\rm eff},X}
=  \left( a^2 \langle \frac{\partial^2 V}{\partial \phi_i^2} \rangle
- \frac{a''}{a} \right)/ [\sqrt{\lambda} \phi_0]^2$ \\
\hline
\end{tabular}
\caption{
%\footnotesize 
Translation table between quantities in plots and superpotential
parameters (we give the eqn. where the parameter is defined). Recall that
$\eta$ is the conformal time coordinate, $ \eta = 0 $ at the end of
inflation, and subsequently the scale factor is  $a( \eta)/a(0) \simeq
\eta/2 + 1$.}
\label{tab2}
\end{table*}
\renewcommand{\arraystretch}{1}
%\end{widetext}
%
For numerical convenience, we consider a chaotic inflationary scenario
with a quartic potential for the inflaton. More specifically, we focus on
the superpotential~\footnote{To embed the system in a supergravity context
while preserving a flat potential for the inflaton field, one may
impose~\cite{natch} a definite parity for the K\"ahler potential ${\cal K}
= {\cal K} \left( \Phi - \Phi^* \right) \,$. Doing so, the inflaton is
identified with the real direction of the scalar component of $\Phi\,$,
and supergravity corrections can be neglected.}
\begin{equation}
W\left(\Phi,
N\right)=\frac{\sqrt{\lambda}}{3}\,\Phi^3+\frac{1}{2}\,
\left(\sqrt{2}\,g\,\Phi+M\right)\,N^2\,\,.
\label{w1}
\end{equation}
The second term of $W$ reproduces the lagrangian~(\ref{lag}) for the r.h.
neutrinos. We denote the scalar components of the inflaton and of the r.h.
neutrinos multiplets with $\phi$ and $N\,$, respectively. To simplify the
numerical computations, the imaginary components of the scalar fields will
be neglected. Therefore, after canonical normalization, $\phi\rightarrow
\phi/\sqrt{2},\,N\rightarrow N/\sqrt{2}\,$, we consider the scalar
potential~\footnote{We neglect the term quartic in $N\,$, subdominant with
respect to the mass term $M^2 \, N^2$ during most of the
preheating/rescattering period, as well as the mixed term $\propto
\sqrt{\lambda} \, g \, \phi^2 \, N^2\,$, negligible with respect to the
mixed term present in eqn.~(\ref{pot1}) for $g^2 \gg \lambda \,$.
One may also be worried that, if the right--handed sneutrino is charged
under some gauge group (as it generally happens in grand unified models)
with a gauge coupling $g_G $ not much smaller than one, the corresponding
$D-$term $\propto g_G^2\,\left| N\right|^4 \subset V \,$ could prevent the
amplification of $N$ at preheating-rescattering. However, at least as long
as $\langle N \rangle$ is smaller than the scale at which the gauge
symmetry is broken, this term gets actually compensated by a shift of the
(much heavier) field responsible for the breaking of the symmetry, and it
is thus absent from the effective potential for $N\,$~\cite{ad,muya}.}
\begin{equation}
V_{\rm
{scalar}}=\frac{\lambda}{4}\,\phi^4+\frac{1}{2}\,
\left(g\,\phi+M\right)^2\,N^2\,\,.
\label{pot1}
\end{equation}
The size of the temperature fluctuations of the Cosmic Microwave
Background sets $\lambda \simeq 9\times 10^{-14}\,$, while the neutrino
mass $M$ as well as the coupling $g$ to the inflaton are model dependent
parameters. The case $M=0$ is analyzed in detail in~\cite{felderkofman}.
\begin{figure}
\epsfig{file=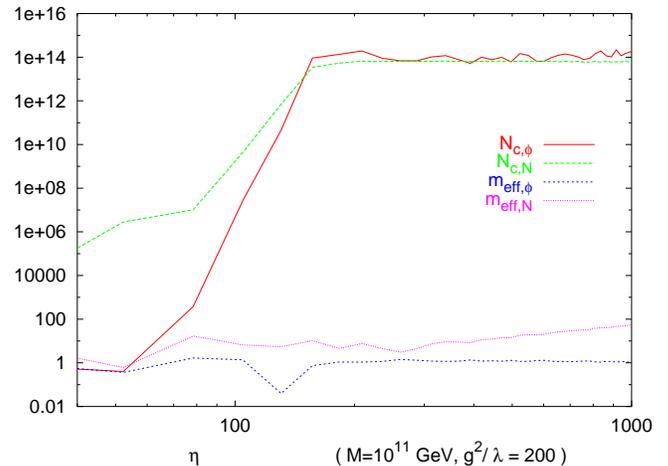,angle=-90,width=.5\textwidth}
\caption{
%\footnotesize 
Time evolution of the comoving number density and of the effective masses of the inflaton $\phi$ and of the r.h. sneutrino $N\,$. See table \ref{tab2} for
notation.}
\label{fig:fig1}
\end{figure}
In figure~\ref{fig:fig1} we show the time evolution of the comoving number
density and  of the comoving effective mass of the two scalars $\phi$ and
$N\,$, for the  particular choice of the parameters $M=10^{11} {\rm
GeV}\,$, ${\tilde g} \equiv g^2 / \lambda = 200\,$. The effective mass is
defined as
\begin{equation}
m_{\rm eff, \phi_i}^2 \equiv a^2 \langle \frac{\partial^2 V}{\partial \phi_i^2} \rangle - \frac{a''}{a} \,\,,
\label{mass}
\end{equation}
where $a$ is the scale factor of the Universe, normalized to one at the
end of inflation, prime indicates derivative with respect to the conformal
time $\eta\,$, while $\langle \dots\rangle$ denotes average over the sites
of the lattice. The term $a''/a$ appears in eqn.~(\ref{mass}) because we
are considering minimally (rather than conformally) coupled scalars, and
it vanishes in a radiation--dominated background. The comoving number
density is  defined as the integral over momentum of the ``occupation
number''
\begin{eqnarray}
n_k \left( \eta \right) &\equiv& \frac{1}{2} \left( \omega_k \, \vert f_k \vert^2 + \frac{1}{\omega_k} \, \vert f_k' \vert^2 \right) \,\,, \nonumber\\
\omega_k^2 &\equiv& k^2 + m_{\rm eff}^2 \,\,,
\label{n}
\end{eqnarray}
where $f_k$ denotes the Fourier transform (to be evaluated on the lattice)
of the rescaled field $a \,\phi\,$. By definition~\cite{latticeeasy}, the
quanta stored in the oscillating inflaton condensate do not contribute to
$N_{c,\phi}$ in figure~\ref{fig:fig1}. The three quantities $m_{\rm
eff}\,,N_{\rm c}\,,$ and $\eta$ are all shown in units of $\sqrt{\lambda}
\, \phi_0 \simeq 1.25 \cdot 10^{12} \, {\rm GeV} \,$, with $\phi_0 \simeq
0.342 \, M_P$ denoting the value of $\phi$ at the end of inflation, to the
appropriate power. All the numerical results presented in this work are
obtained with a two dimensional lattice of size $L = 20 \left(
\sqrt{\lambda} \, \phi_0 \right)^{-\,1}$ and with $N=1024^2$ sites
(see~\cite{latticeeasy} for details).
Figure~\ref{fig:fig1} exhibits the features that we have outlined at the
beginning of this section, namely a quick stage of exponential growth of
the occupation numbers followed by a period in which the occupation
numbers are nearly constant. During the first stages of the process, the
results presented reproduce very well the ones obtained
in~\cite{felderkofman} for $M=0\,$, since the ``bare'' mass of the r.h.
neutrinos is initially negligible.
However, the presence of a non vanishing bare mass affects the subsequent evolution of the system. Indeed, when the value of the fluctuations in the sneutrino field becomes comparable with the amplitude of inflaton oscillations, the second term in eqn.~(\ref{pot1}) shifts the minimum of the effective potential of the inflaton, giving it an effective mass that is roughly constant in comoving units. A stronger effect is related to the fact that the sneutrino itself is massive. In rescaled units and for the present choice of the parameters, we have
\begin{equation}
{m_{\rm eff, N}}^2 \simeq \left\langle \left( 0.08 \, a + 14.1 \, \frac{a \, \phi}{\phi_0} \right)^2\right\rangle \,\,.
\label{mn}
\end{equation}
Numerical results show that $\langle \phi \rangle \sim \phi_0/a\,$ (as one
could also see by inspecting the potential~(\ref{pot1}) in Hartree
approximation), so that the part in the above equation that depends on
$\phi$ remains of order one. It follows that the  r.h. neutrino mass $M$
dominates the effective mass~(\ref{mn}) for $\eta \geq {\hat \eta} \sim
350\,$, as clearly indicated by the growth of $m_{\rm eff, N}$ visible in
figure~\ref{fig:fig1} for $\eta > {\hat \eta}\,$.
Production of sneutrinos at preheating in the present model is strictly related to the production of fermions we have analysed in section 2.3~. In particular, production occurs whenever the effective neutrino mass (\ref{mn}) crosses zero. Numerical results show that preheating is terminated by rescattering effects when the scale factor $a$ is of the order of $a_{\rm {resc}}\simeq 100$. As a consequence, sneutrinos with a bare mass up to $g\,\phi_0/a_{\rm {resc}}\simeq g\cdot 10^{17}\,$GeV will be efficiently produced at preheating and will constitute a sizable fraction of the background energy \footnote{Non-adiabatic production of sneutrinos can occur for bare masses as large as $g\,\phi_0/4\sim g\cdot 10^{18}\,$GeV, but the efficiency of the process will be much lower, because redshift effects will terminate the resonance before rescattering sets in.}.
\begin{figure}
\psfig{file=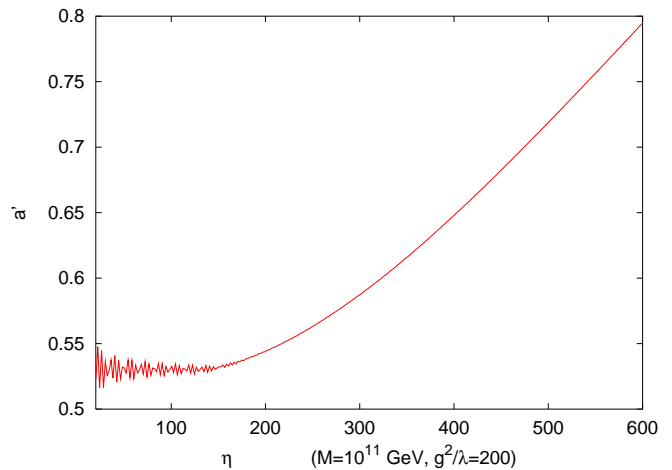,angle=-90,width=.5\textwidth}
\caption{
%\footnotesize  
Time evolution of the derivative of the scale factor with respect to conformal time. A constant value indicates radiation domination, while a linear growth indicates matter domination.}
\label{fig:fig2}
\end{figure}
Numerical results show that after the onset of rescattering, the energy
density gets roughly equiparated between the quanta of the two species. As
a consequence, in general large couplings correspond to a large
interaction energy, and therefore to a smaller number density during the
rescattering/thermalization stage~\cite{cldeca,felderkofman}. In
particular, for this reason a large quartic self--coupling $g_{\rm
N}^2\,\left|N\right|^4$ for the sneutrino would prevent it from getting
large occupation numbers, since energy conservation would impose $\langle
N^2\rangle\propto  \sqrt{\lambda/g_{\rm N}^2}$. Numerical results also
indicate that soon after the beginning of rescattering most of the r.h.
neutrino quanta have a momentum of the order $k_* \sim 15 \,
\sqrt{\lambda} \, \phi_0\,$.~\footnote{The precise value of the typical
momentum $k_*$ of the distributions, as well as the time needed for $N_c$
to saturate, are a nontrivial function of ${\tilde g}\,$, since different
values of this parameter lead to different positions (in momentum space)
and strengths of the resonance bands~\cite{nonthana2}. However, the
rescattering stage destroys these resonance bands, making the dependence
of $k_*$ on ${\tilde g}\,$ milder.} Thus, most of the r.h. neutrinos
become non relativistic at a time not much greater than ${\hat \eta}$.
From this time on, the energy density of the system redshifts as the
energy density of matter. The transition  between the two stages of matter
and radiation domination is clearly visible in figure~\ref{fig:fig2},
where we show the time evolution of the derivative of the scale factor
with respect to conformal time. As long as the neutrino mass is
negligible, the energy density of the system redshifts as the one of
radiation~\cite{nonthana2,felderkofman}, and the evolution of the scale
factor is very well approximated by $a \simeq \left( 1 + t \right)^{1/2}
\simeq \eta/2+1\,$, where we have set $t=\eta=0\,$ at the end of
inflation. Therefore, during the initial stage of radiation domination,
$a'$ is constant. In the following matter dominated stage $a \propto
\eta^2\,$, and $a'$ grows linearly with time.
To estimate the baryon asymmetry produced from the decay of the r.h.
sneutrinos, we need to know the fraction of the entropy of the Universe
that is generated in the decay. The baryon asymmetry will be~\cite{HMY}
\begin{eqnarray}
Y_B &\simeq &  -\epsilon \frac{8}{23} \frac{N_{N}}{s_{N}}\frac{s_{N}}{s_{tot}}
 \simeq \frac{8}{23}\,\left(- \epsilon\,\frac{3}{4}\,\frac{T_N}{M}\right) \frac{s_{N}}{s_{tot}}   \nonumber\\
&\simeq& 0.3\times 10^{-10} \left(\frac{T_N}{10^6\,{\rm
{GeV}}}\right)\,\left(\frac{m_3}{0.05\,{\rm {eV}}}\right)\,\frac{s_{N}}{s_{tot}} \delta_{\rm
CP} \,\,,
\label{asdom}
\end{eqnarray}
where $s_{N} \propto g_*(T_{N})T_{N}^3$ is  the entropy produced in $N$ 
decay, and $s_{tot}$ is the total entropy of the Universe. If the r.h.
sneutrinos dominate the Universe when they decay, then $T_{N}$ is  the
temperature to which the Universe is reheated by the decay of the
sneutrinos $N$, and $s_{tot} \sim s_{N}$. This condition is satisfied if
the inflaton mainly decays only into one r.h. sneutrino family (as it is
clearly the case in the one generation model we have studied numerically).
However, if the inflaton couples to other scalars (also in particular to
the other generations of sneutrinos), these could produce additional
entropy.
At variance with the case of leptogenesis induced by the decay of
right--handed neutrinos, analysed in section 2.3, the baryon
asymmetry~(\ref{asdom}) does not depend on the r.h. (s)neutrino mass, that
must only satisfy $M>T_N$ in order to prevent thermal regeneration of the
r.h. sneutrinos after their decay. This is due the fact that, thanks to
Bose statistics, r.h. sneutrinos can get large occupation numbers at
preheating (whereas Pauli blocking makes fermion production less
efficient), and they can easily represent a substantial fraction of the
energy in the Universe.~\footnote{Notice that for a massless inflaton, the
inflaton energy redshifts as radiation, and non-relativistic neutrinos
will easily dominate the energy in the Universe. If the inflaton in
instead massive, then the energy in sneutrinos would in any case be a
fraction of order unity of the background energy, and would start
increasing after inflaton decay. As a consequence, the resulting baryon
asymmetry would still be given (at most up to factors of order one) by the
expression~(\ref{asdom}).}
The generalization to the more realistic case of three neutrino families coupled to the inflaton is obtained by considering the following superpotential, in the mass eigenstate basis for the r.h. neutrinos,
\beq
W = \frac{\sqrt{\lambda}}{3} \Phi^3 + h_{ji} \,L_i\cdot H_u \, N_j+\frac{1}{2}\,(M_j \delta_{jk} + \sqrt{2} g_{jk} \Phi)N_k\, N_j.
\eeq
This gives a potential for the real components of the scalars
\begin{equation}
V= \frac{{\lambda}}{4} \phi^4 + \frac{1}{2}\sum_{i,j,k} (g_{ij} \phi + M_i \delta_{ij}) (g_{ik}^* \phi + M_i \delta_{ik})N_j N_k + ... 
\end{equation}
where dots include the terms involving the Yukawa $h$, which are not
relevant for nonthermal $N_i$ generation. We suppose for simplicity that
the neutrino--inflaton coupling $g_{ij}$ is diagonal. Energy
considerations after rescattering~\cite{cldeca,felderkofman} lead to the
expectation that the sneutrino family that is most strongly coupled to the
inflaton is also the one that will have the smallest number density
(clearly, on the assumption that all the sneutrinos are sufficiently
coupled to be amplified). This is opposite to the scenario in which
sneutrinos are produced by the perturbative decay of $\phi$, where
$N_{N_i}$ is proportional to the inflaton branching ratio to $N_i$.
However, the presence of the r.h. sneutrino bare masses, as well as the
existence of couplings to the Standard Model degrees of freedom (whose
effects will be analysed in detail in the next sections), can strongly
affect these conclusions.
Although the resulting baryon asymmetry~(\ref{asdom}) has the same
expression as the one reported in~\cite{HMY}, the mechanism that led to a
sneutrino dominated Universe is different from the generation of large
expectation values considered in ref.~\cite{HMY} or in the
Affleck--Dine~\cite{ad} mechanism. Indeed, the latter is effective if
during inflation the sneutrino (or, more generally, the amplified flat
direction) has a mass much smaller than the Hubble rate. This requires
(besides a sufficiently small bare mass $M\ll H$) that the amplified field
does not get a large effective mass through its coupling to the inflaton.
It is important to remark that the mechanism we are discussing can provide
a sufficient leptogenesis even if the coupling $g$ between the inflaton
and the r.h. neutrino multiplets is much smaller than the one needed in
non supersymmetric models, i.e. with only the production of the neutrinos
taken into account, see eqn.~(\ref{asinfdom}). Anyhow, even couplings as
small as $g^2 \sim 10 \, \lambda$ prevent the formation of a large
condensate during inflation. Therefore, the two mechanisms can lead to
large occupation numbers in complementary regions of the parameter space.
Notice that the above discussion applies to every effective mass term that
can arise in the potential for a (quasi) flat direction. In particular, it
could be interesting to consider the effective mass of the order of the
Hubble parameter that is generally induced by supergravity
corrections~\cite{DRT}, although in this case amplification effects may be
weakened by the quick redshift characterizing the nonrenormalizable
interactions. If supergravity corrections induce a tachyonic mass
$m^2_{\rm {eff}}\simeq -H^2$, a large expectation value will be generated
during inflation~\cite{DRT}, and the dynamics of preheating will turn out
to be rather different from the one considered in the present section.
This however requires a suitable nonminimal K\"ahler potential, and we
will not discuss this possibility in this work.
Finally, it is worth stressing that both the leptogenesis scenario
described in this section and the one considered in~\cite{HMY}, although
they are related to Affleck--Dine leptogenesis, are somehow different from
it for what concerns the fulfillment of the Sakharov $CP$--violation
condition~\cite{Sakharov:1967dj}.  In the Affleck--Dine scenario, the
latter is achieved by the motion of the Affleck--Dine condensate (that
requires coherence over many Hubble lengths), while in the mechanism we
are analysing, $CP$--violating sneutrino decays are crucial in the
generation of an asymmetry.
\section{Production of light degrees of freedom at rescattering}
The description presented above is further modified by the effects of the
coupling of the r.h. neutrino multiplets to the l.h. leptons and
Higgses, coming from the superpotential~(\ref{superpotential}).
Considering for simplicity only one generation, the
superpotential~(\ref{w1}) will be supplemented by
\begin{equation}
\Delta W = h \, N \, L \cdot H \,\,.
\label{delw}
\end{equation}
The resulting scalar potential contains several interaction vertices
coming from F--terms. In addition, there are quartic contributions from the
D--terms. The presence of the latter in the potential plays an important
role for our discussion, since rescattering mostly affects the D--flat
directions. To see this, consider the  case in which the left--handed
selectron and the charged scalar Higgs vanish. The $D$-terms for the
neutral Higgs and sneutrino then take the familiar form of the tree level
MSSM Higgs potential, with $H_d^0$ replaced by $\tilde{\nu}_L$
\begin{equation}
V_D=\frac{g_{{\rm SU} \left( 2 \right)}^2 + g_Y^2}{8}\,  \left\vert \left\vert H^0
\right\vert^2-\left\vert \tilde{\nu}_L \right\vert^2\right\vert^2\, ~~. 
\end{equation}
The directions that are not D--flat (i.e. the ones for which $\vert
\tilde{\nu}_L \vert \neq \vert H^0\vert$ in the present example) are
characterized by a large (gauge) quartic coupling in the scalar potential.
Due to this coupling, they cannot be significantly amplified by
rescattering effects. On the contrary, D--flat directions have only
quartic couplings coming from F--terms as $\vert \partial W / \partial N
\vert^2\,$, whose strength $h^2$ will be typically taken $\ll 1$ in all
the cases considered below. Indeed these quartic interactions among the
D--flat directions will be neglected in our computation, since they can be
relevant only during the thermalization stage, when the variances of these
fields have grown to be sufficiently large.
The most important F--terms arising from the total superpotential are of
the form $\sim h^2\,\vert N\vert^2\left(\vert \tilde{\nu}_L\vert^2+\vert
H^0\vert^2\right)\,$. If we denote by $X$ the D--flat combination $\vert
\tilde{\nu}_L\vert=\vert H^0\vert$, then the relevant coupling of $X$ to
the sneutrino field will be simply given by $h^2\,N^2\,X^2\,$ (as in the
previous section, we consider for numerical convenience only real
directions). Besides the quartic term $\propto h^2 \, X^4\,$, we will also
neglect the interaction term $\propto \left( g \phi + M \right) N \, X^2
\simeq M \, N\, X^2\,$, which is responsible for the late time decay of
the r.h. sneutrinos (that is, the supersymmetric counterpart of the vertex
which gives the decay of the r.h. neutrinos into Higgses and leptons).
We thus consider the simplified model characterized by the scalar potential
\begin{equation}
V\left(\phi,\,N,\,X\right)=\frac{\lambda}{4}\,\phi^4+\frac{1}{2}
\left(M+g\,\phi\right)^2\,N^2+\frac{1}{2}\,h^2\,N^2\,X^2\,\,.
\label{simple}
\end{equation}
Neglecting the imaginary directions of the scalar fields, as well as many
of the interaction terms, allows a considerable reduction of the computing
time needed for the numerical simulations (this is particularly welcome
for the extensive computation that we discuss in the next section). The
above discussion leads us however to believe that the simplified model
should well describe the main features of the preheating  and rescattering
process for the supersymmetrized see-saw model with a nonperturbative
production of the r.h. neutrinos.
\begin{figure}
\psfig{file=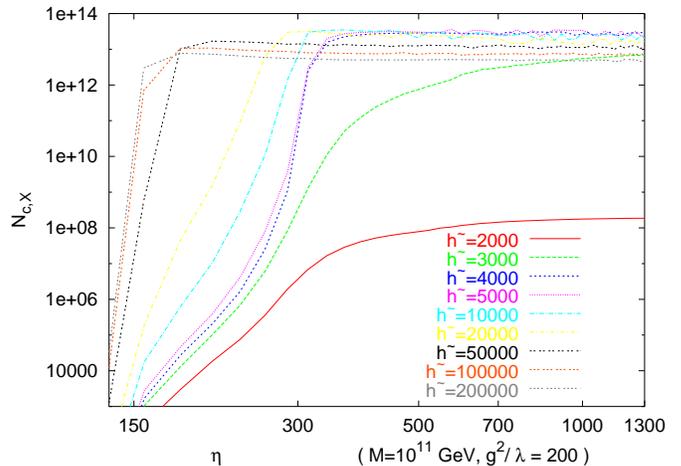,angle=-90,width=.5\textwidth}
\caption{
%\footnotesize 
Time evolution of the comoving number density of the light quanta $X\,$.
See table \ref{tab2} for
notation.}
\label{fig:fig3}
\end{figure}
In figure~\ref{fig:fig3} we show the time evolution of the comoving number
density of the quanta of $X\,$. As in the previous section, we have fixed
$M=10^{11}\,$ GeV, ${\tilde g}=200\,$, while different values of the
parameter ${\tilde h} \equiv h^2/\lambda$ are shown. Even if in the
simplified model~(\ref{simple}) the $X$ field is not directly coupled to
the inflaton, we see that (for suitable values of the coupling $h$) it can
be highly amplified by the rescattering of r.h. sneutrino quanta.
Figure~\ref{fig:fig3} shows that the growth of number of $X$ particles is 
roughly exponential in time. When the effective sneutrino mass is varying 
non-adiabatically in time and is not negligible with respect to sneutrino
typical momenta, the production of $X$ particles cannot be analysed in
terms of scatterings of sneutrinos. However, after the end of the
parametric resonance period and the onset of rescattering,  one can expect
that a particle--like picture can give some description of the behavior of
the system~\cite{felderkofman}. In this case, if the dominant contribution
to $X$ production process were given by the $2\rightarrow 2$ scattering
$NN\rightarrow XX$, the rate of growth of $N_X$ should be proportional to
$h^4$. The lattice results appear to present a milder dependence on $h$,
suggesting that the $NN\rightarrow XX$ scatterings alone cannot account
for the production of $X$ states. In a naive perturbative analysis, the
contribution,  from  $(m+2)\times N\rightarrow XX$ processes, to the rate
of growth of $N_X$,  is proportional to the  $m\times N\rightarrow XX$
rate times a factor roughly given by $h^4\,N_N^2/(4 \pi p^6) \,$, where $p
\sim 15 \sqrt{\lambda} \phi_0/a$ is the typical momentum exchanged. Due to
the high density of sneutrinos, the expansion parameter $h^4\,N_N^2/(4 \pi
p^6)$ is of order unity  for the values of $h$ we are considering.
Therefore, strongly turbulent processes with many incoming sneutrinos can
contribute significantly to the rate of growth of $N_X$. This is confirmed
by the fact that the total number of particles decreases during the stage
of generation of the $X$ states, thus showing that particle fusion
processes are dominant at this stage~\cite{felderkofman}.
\begin{figure}
\psfig{file=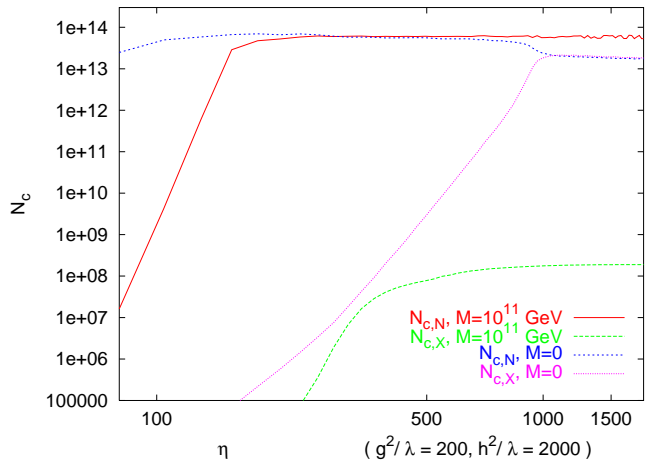,angle=-90,width=.5\textwidth}
\caption{
%\footnotesize 
Comparison of the growth of the occupation numbers of $N$ and $X$
with and without the r.h. neutrino mass term. Notice that the
amplification of $X$ considerably weakens when the r.h. neutrino quanta
become non-relativistic, $\eta \sim 350\,$.See table \ref{tab2} for
notation.}
\label{fig:fig4}
\end{figure}
The main features shown in figure~\ref{fig:fig3} are shared by the other
evolutions with different ${\tilde g}$ that we will consider in the next
section, and they can be understood at least qualitatively. As could have
been easily guessed, the timescale for the growth of $N_{c,X}$ is a
decreasing function of ${\tilde h}\,$. We also notice from the figure that
the amplification of $X$ becomes less efficient as the quanta of $N$
become non relativistic, at $\eta \sim 300\,$. This can be seen explicitly
in figure~\ref{fig:fig4}, where we show the effect of the r.h. neutrino
mass term on the growth of the comoving occupation numbers of the $N$ and
$X$ fields. If the two fields are both massless, the rescattering of the
quanta of $N$ lifts the $X$ to (practically) the same
amplification~\cite{felderkofman}. We observe that the situation is
completely different for the case in which the quanta of $N$ have a
sufficiently high mass. Indeed, when the amplification of $X$ from the
r.h. neutrino quanta substantially decreases when the latter become
non-relativistic. As a consequence of the two effects mentioned in this
paragraph, the asymptotic value of $N_{c,X}$ (at least in the time
interval we have considered) decreases at small ${\tilde h}\,$.
Figure~\ref{fig:fig3} also shows a decrease of the asymptotic $N_{c,X}$
for high ${\tilde h}\,$. As discussed in~\cite{felderkofman}, for
sufficiently high coupling $h$ the two fields have comparable occupation
numbers, $N_{c,N} \simeq N_{c,X}\,$. Clearly, the higher the coupling is,
the sooner this (approximate) equipartition is reached. For a very high
$h\,$, the potential energy associated to the last term of~(\ref{simple})
then disfavors the production of the quanta of the two fields, in the same
way as a high quartic coupling $\propto N^4\,$ added to the
potential~(\ref{pot1}) would have prevented the amplification of the r.h.
neutrinos in the two fields case.
\begin{figure}
\psfig{file=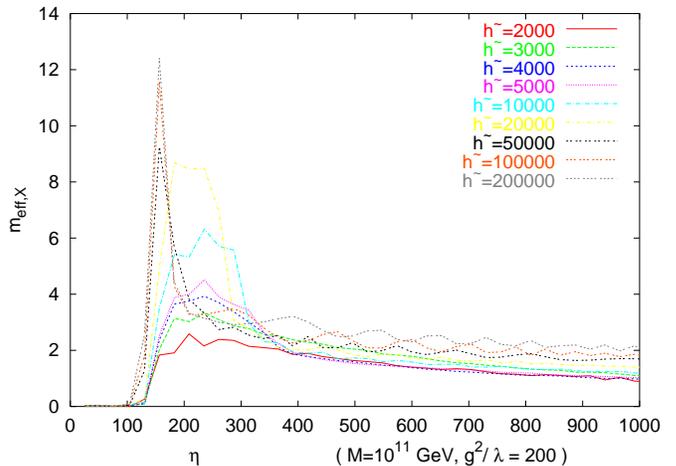,angle=-90,width=.5\textwidth}
\caption{
%\footnotesize 
Time evolution of the effective mass of the quanta of $X\,$.
See table \ref{tab2} for notation.}
\label{fig:fig5}
\end{figure}
In figure~\ref{fig:fig5}, we show the time evolution of the effective mass
of the quanta of $X\,$. Comparing it with their distribution in momentum
space, one realizes that most of the quanta are always in a relativistic
regime.
\section{Perturbative production of gravitinos}
The light quanta $X$ generated at rescattering can in turn be responsible
for the production of unwanted relics such  as gravitinos. If unstable,
gravitinos with a mass of the order of the electroweak scale (which is
what we expect in models of gravitationally mediated supersymmetry
breaking) disrupt the successful predictions of primordial
nucleosynthesis, unless their abundance is below the very stringent
bound~\cite{grnuc}
\begin{equation}
Y_{3/2} \la 10^{-14} \left( {\rm TeV} / m_{3/2} \right) \,\,.
\label{gralim}
\end{equation}
In inflationary theories, several sources of gravitino production have
been considered. The most standard of them is the perturbative production
from the thermal bath formed at reheating. In this case, the above
limit~(\ref{gralim}) translates into the upper bound $T_{\rm rh} \la {\rm
few} \times 10^{10} \,$ GeV on the reheating temperature~\cite{grath}.
Other sources of gravitino production can be studied. Since gravitinos
arise in supersymmetric models, a very natural ``candidate'' channel for
their production is the decay of the inflaton into its supersymmetric
partner (the inflatino) plus a gravitino. This process is however either
kinematically forbidden~\cite{grreh} or strongly suppressed~\cite{nop} by
the fact that the difference between the inflaton and the inflatino mass
is governed by the supersymmetry breaking scale, which is of the same
order of the gravitino mass. The resulting gravitino production is
sufficiently small. Recently, the generation of gravitinos at preheating
has been extensively discussed, both concerning the relatively easier
case of the transverse component~\cite{mama}, and the more delicate issue
of the longitudinal component~\cite{granonth1,granonth2,granonth3}. These
studies are focused on the nonperturbative amplification of the gravitino
field due to the coherent oscillations of the inflaton, and this mechanism
of production is found to be sufficiently limited~\cite{granonth2}
provided that the inflationary sector of the theory is weakly (e.g.
gravitationally) coupled to the one responsible for the present
supersymmetry breakdown.
In this section we discuss a different possible source of gravitinos,
namely a perturbative production from the nonthermal distributions of
light MSSM quanta generated at rescattering. A comparison with the
standard thermal production may be used as an initial motivation.
Concerning the latter, the requirement of a low reheating temperature can
be seen as the demand that the inflaton decays sufficiently late, so that
particle in the thermal bath have sufficiently low number densities and
energies when they form. If $H \simeq 10^{12} \,$ GeV at the end of
inflation, and if the scale factor $a$ is normalized to one at this time,
the generation of the thermal bath cannot occur before $a \simeq 10^7\,$.
Gravitino overproduction is avoided by the fact that in the earlier times
most of the energy density of the Universe is still stored in the coherent
inflaton oscillations. As we have seen in the previous sections, this last
assumption is no longer valid if preheating and rescattering effects
are important. Indeed, in the model considered above the energy density of
the scalar distributions becomes dominant already when the scale factor is
of order $100$ (the precise value being a function of the parameters of
the model).~\footnote{The situation is even more enhanced for hybrid
inflationary model, in which the energy density of the zero modes of the
scalars gets dissipated within their first
oscillation~\cite{tac,abc,cpr}.}
Although this comparison is rather suggestive, it is fair to say that the
computation of gravitino production in the context of rescattering is
certainly more difficult than the usual thermal production. While in the
latter case a perturbative approach can be adopted, and the final result
can be readily estimated by computing the rate of $2 \rightarrow 2$
processes with one gravitino in the final state, rescattering is a highly
nonlinear phenomenon. In the bosonic case, we already remarked that naive
perturbative  estimates poorly reproduce the initial amplification of the
fields $X\,$. Only towards the end of the rescattering stage the number
densities of the amplified fields become sufficiently small so that $2
\rightarrow 2$ processes become dominant, as the (approximate)
conservation of the total comoving occupation number at relatively late
times signals~\cite{felderkofman}. Unluckily, Pauli blocking forbids
fermionic fields to behave classically (in the sense discussed in section
$3$), and lattice simulations cannot be used. However, we may take the
numerical results for bosons as a guideline. Also for the production of
fermions  more complicated processes than just $2 \rightarrow 2$
interactions could be relevant during most of the rescattering stage,
while they should be subdominant at sufficiently late time. The latter is
presumably set by the same timescale at which rescattering is seen to end
in the numerical simulations described in the previous sections. With this
in mind, we proceed to an estimate of the amount of gravitinos produced by
$2 \rightarrow 2$ processes with the fields amplified at
preheating/rescattering in the incoming state. We stress once more that
this estimate can be reliable only from the beginning of the
thermalization stage on, so it should provide  a lower bound on the total
number of produced gravitinos. It is possible that a higher amount of
gravitinos is produced at earlier times, when nonlinear effects cannot be
neglected.
This computation has been carried out in appendix, where also some details
(i.e. concerning the quantization of the system and the choice to focus on
the transverse gravitino component) are reported. As for the thermal
production, the dominant $2 \rightarrow 2$ processes have only one
gravitino in the final state, and hence only one gravitationally
suppressed vertex. In the thermal case, the dominant processes have a
gauge interaction as the second vertex, consider e.g. the process $X \, X
\rightarrow {\tilde z} \, \psi_{3/2} \,$ with one higgsino in the
propagator. For the nonthermal distributions of scalars that we are
considering, however, such processes are kinematically forbidden. This is
a crucial point, which poses a significant limit on the estimated
production of gravitinos. We can easily understand it using the specific
process just mentioned as an example: either the Higgses are not amplified
at rescattering (so the above process is irrelevant) or the non-vanishing
$\langle H^2 \rangle$ provides an effective mass to the zino produced in
the interaction. When the light scalar distributions saturate (that is,
when the gravitino production can be effective), we find numerically $a \,
\sqrt{\langle X^2 \rangle} \sim \left( 10^{-2} - 10^{-1} \right) \phi_0
\,$. The typical comoving momenta $a\,p$ characterizing the scalar
distributions are instead only about one order of magnitude higher than
the ``inflaton mass'' at the end of inflation, $a \, p \sim 15
\sqrt{\lambda} \, \phi_0 \,$. As a consequence, the gauginos acquire a
mass $m_{\tilde z} \sim \left( 10^2 - 10^3 \right) p\,$, which shows that
these processes are kinematically forbidden.~\footnote{Processes with an
additional gauge interaction and the gaugino in the propagator are allowed
but strongly suppressed. See the appendix for details.}
We notice that, at least for this specific kind of interactions, the
system is still effectively behaving as a condensate: the number densities
of the scalar distributions are set by the quantity $\sqrt{\langle X^2
\rangle}\,$, which is much higher than the typical momenta of the
distributions themselves. This generates a high effective mass for all the
particles ``strongly'' coupled to these scalar fields. Compare this
situation with a medium in thermodynamical equilibrium: in this case both
the typical momenta and the effective masses are set by the only energy
scale present, namely the temperature of the system. As should be clear by
the above discussion (see also~\cite{felderkofman}), the thermalization of
the distributions produced at rescattering necessarily proceeds through
particle fusion. Only after a sufficiently prolonged stage of
thermalization, will  the system be sufficiently close to thermodynamical
equilibrium so to render processes as the one discussed above
kinematically allowed.
As we discuss in the appendix, kinematically allowed $2 \rightarrow 2$
interactions can be obtained by taking a trilinear interaction coming from
the superpotential term (\ref{delw}), also responsible for the Dirac mass
term for the neutrinos. This can lead to processes of the kind ${\tilde
N}_R \,  X \rightarrow x \, \psi_{3/2}\, $, or $X \, X \rightarrow N_R \,
\psi_{3/2}\,$. The physical number density of (transverse) gravitinos
produced by these processes can be estimated as
%\begin{widetext}
\begin{equation}
N_{3/2} \left( \eta \right) \sim \frac{h^2}{M_P^2} \, \left[ \frac{N_{c,X}  \, N_{c,N}}{a^6} \right]_{\Big \vert_{\eta=\eta_*}} \!\! \left[ \frac{a \left( \eta_* \right)}{a \left( \eta \right)} \right]^3 \;\;\;,\;\;\; \eta > \eta_* \,\,,
\label{granum3}
\end{equation}
%\end{widetext}
%
where $\eta_*$ is the time (to be determined numerically) at which the
expression in the first parenthesis reaches its maximum, while the second
parenthesis is a dilution factor due to the expansion of the Universe at
later times (see the appendix for details).~\footnote{The time $\eta_*$
roughly corresponds at the moment at which the distribution of light
quanta $X$ starts to saturate. This typically occurs towards the end of
the rescattering stage, which guarantees that considering only $2
\rightarrow 2$ processes should provide at least an order of magnitude
estimate of the gravitino quanta produced at this stage.}
As we have remarked, this result is subject to the limit~(\ref{gralim}),
where $Y_{3/2} \equiv N_{3/2}/s\,$, and $s$ is the entropy density of the
Universe, computed once the dominating thermal bath is formed. For
practical use, we find that a more ``convenient'' bound can be obtained if
(\ref{gralim}) is combined with the result for the baryon asymmetry,
eqn.~(\ref{asdom}). For this purpose,  consider the ratio
\begin{equation}
\zeta \equiv \frac{Y_{3/2}}{Y_B} = \frac{23}{8\,\epsilon_1} \, \frac{N_{3/2} \left( \eta \right)}{N_N \left( \eta \right)} = \frac{23}{8\,\epsilon_1} \, \frac{N_{3/2} \left( \eta_* \right)}{N_N \left( \eta_* \right)} \,\,.
\end{equation}
The quantity $\zeta$ has two main advantages, (i) it is independent of the
entropy of the Universe and (ii) it can be computed already at  $\eta =
\eta_*\,$, since after this time the two physical number densities
$N_{3/2}$ and $N_N$ simply rescale as $a^{-3}\,$. While $Y_{3/2}$ must
satisfy the upper bound~(\ref{gralim}), the limit $Y_B \ga 10^{-11}$ poses
a lower bound on the number density of the sneutrinos, if leptogenesis is
assumed to be responsible for the baryon asymmetry of the Universe.
Adopting the parameterization~(\ref{param}), we then see that the ratio
$\zeta$ has to satisfy
\begin{equation}
\zeta \simeq \frac{3 \times 10^6}{\delta_{\rm CP}} \, \left( \frac{10^{10} \, {\rm GeV}}{M} \right) \, \left( \frac{0.05 \, {\rm eV}}{m_3} \right) \, \frac{N_{3/2} \left( \eta_* \right)}{N_N \left( \eta_* \right)} < 10^{-3} \,\,.
\label{finalbound}
\end{equation}
It is important to stress that, unlike the limit~(\ref{gralim}), this
bound cannot be ameliorated by an eventual entropy release which may
occur between the decay of the r.h. sneutrinos and nucleosynthesis, since
both the gravitino and the baryon number densities would be diluted in the
same amount. For this reason, we find in the present context the
bound~(\ref{finalbound}) more significant than the limit~(\ref{gralim})
involving the gravitino abundance alone.
\begin{figure}
\epsfig{file=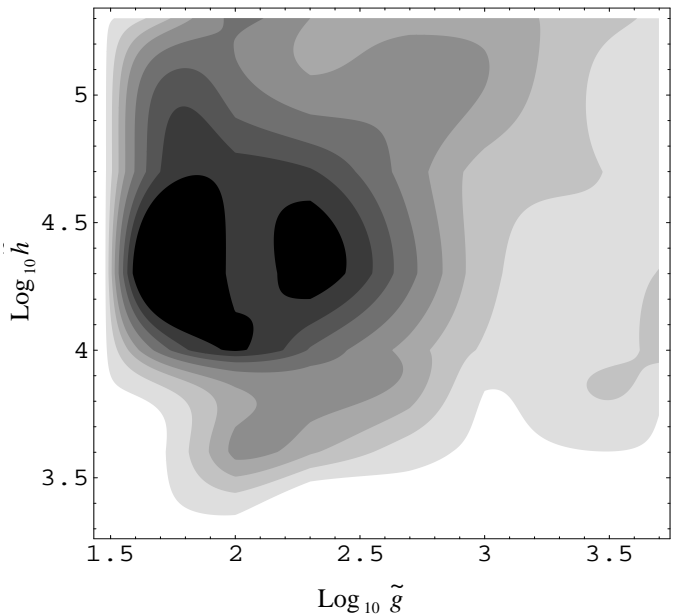,width=.5\textwidth}
\caption{
%\footnotesize  
Contour plot for the quantity ${\tilde \zeta}/{\tilde h}\,$, where ${\tilde \zeta}$ is defined in eqn.~(\ref{zetati}). The contour lines range from ${\tilde \zeta}/{\tilde h} = 3 \cdot 10^{-12}$ to ${\tilde \zeta}/{\tilde h} = 1.1 \cdot 10^{-10}\,$ (darker region).}
\label{fig:fig6}
\end{figure}
We wanted to verify whether the condition~(\ref{finalbound}) is respected
for the choice $M = 10^{11} \, {\rm GeV}$ considered in the previous
sections, and for several values of the couplings $g$ and $h$ (defined in
the potential~(\ref{simple})) in the range ${\tilde g} \, \in \, \left[ 30
, 5000 \right] \;,\; {\tilde h} \, \in \, \left[ 2000 , 200000 \right] \,$
(we remind that ${\tilde g} \equiv g^2 / \lambda$, and analogously for
$h\,$). To do so, we have defined
\begin{equation}
\zeta \equiv \frac{1}{\delta_{\rm CP}} \, \left( \frac{0.05 \, {\rm eV}}{m_3} \right) \, {\tilde \zeta} \,\,,
\label{zetati}
\end{equation}
and in figure~\ref{fig:fig6} we have plotted the quantity ${\tilde \zeta}
/ {\tilde h}$. In this way, we factor out the explicit dependence of
$\zeta$ on ${\tilde h}$ coming from the cross section of the dominant $2
\rightarrow 2$ processes, see e.g. eqn.~(\ref{granum3}).
The qualitative behavior of ${\tilde \zeta} / {\tilde h}$ with $h$
(vertical axis) is easily understood in terms of the arguments used to
explain the results of figure~\ref{fig:fig3}. For relatively low $h\,$,
the amplification of the $X$ field is weak, and so few gravitinos quanta
are produced. As $h$ increases, the amplification becomes stronger, both
regarding the final value at which $N_X$ saturates and the rapidity at
which the saturation occurs. As a consequence, the number density of
produced gravitinos also increases. As $h$ further increases, the rapidity
at which $X$ saturates keeps increasing (and so, the time $\eta_*$ at
which most of the gravitino quanta are produced decreases); however, the
final value of $N_X$ starts to become smaller, leading to the decrease of
${\tilde \zeta} / {\tilde h}$ that we observe in the figure. For fixed
values of $h$, the final result is also first increasing and then
decreasing with $g\,$. This behavior is presumably due to the dependence
on $g$ of the total number of quanta of both the $N$ and $X$ fields
produced at preheating/rescattering (notice that it vanishes both at very
small and very high $g\,$). However, the interpretation is in this case
less clear.
Concerning the value of ${\tilde \zeta}$ itself, in the range of coupling
considered it reaches the maximal value at ${\tilde g} \simeq 100 \;,
{\tilde h} \simeq 200000\,$, where it evaluates to ${\tilde \zeta} \simeq
10^{-5}\,$. From what we have just said, we expect that ${\tilde \zeta}$
starts decreasing at higher $h\,$, although the numerical simulations we
have performed show that the decrease starts only at the highest value of
$h$ that we have considered. From the definition~(\ref{zetati}), we see
that the bound $\zeta < 10^{-3}$ is respected, provided the $CP$ violation
encoded in the parameter $\delta_{\rm CP}\,$ (see eqs.~(\ref{param}) and
(\ref{bound})) is not too small. However, we remind that our estimates
take into account only the gravitino quanta produced from the end of the
rescattering stage on, while a higher production at earlier times cannot
be excluded.
\acknowledgments
The numerical results presented in this paper have been obtained using the
code ``LATTICEEASY''~\cite{latticeeasy}, by G.~Felder and I.~Tkachev. We
acknowledge fruitful discussions with Patrick B. Greene, Lev Kofman,
Gianpiero Mangano, and Michael Pl\"umacher. L.B. thanks the CERN-TH
division where part of this work was done. M.P. thanks the Institute for
Particle Physics Phenomenology at the University of Durham for their
friendly hospitality during the initial stages of this work. The work of
L.B., M.P. and L.S. is supported by the European Community's Human
Potential Programme under contract HPRN-CT-2000-00152 Supersymmetry and
the Early Universe.  The work of L.B. was also supported in part by the
Italian INFN under the project ``Fisica Astroparticellare''. M.P. also
acknowledges support under contracts HPRN-CT-2000-00131 Quantum Spacetime
and HPRN-CT-2000-00148 Physics Across the Present Energy Frontier.
\vspace{2cm}
\renewcommand{\theequation}{A-\arabic{equation}}
\setcounter{equation}{0}
\appendix
\section*{Appendix}
In this appendix we derive eqn.~(\ref{granum3}) of the main text. We first
estimate the number density of gravitinos produced by the nonthermal
distributions formed at rescattering. We remark that at this stage
supersymmetry breaking is controlled by the energy density of these
distributions. The longitudinal gravitino component (i.e. the goldstino,
which is ``eaten'' in the unitary gauge) is thus provided by a linear
combination  of the fermionic superpartners of these scalars, and does not
coincide with the longitudinal gravitino at late times (at least for the
standard case of a present gravitationally mediated supersymmetry
breakdown). For this reason, our discussion will be limited to the
transverse gravitino component $\psi_{3/2}\,$. Its mass is given
by~\cite{moroi}
\begin{equation}
m_{3/2} = {\rm e}^{{\cal K} \left( \phi_i \right)/M_P^2} \, \frac{\vert W \left( \phi_i \right) \vert}{M_P^2} \,\,,
\label{mgra}
\end{equation}
where, we remind, $W$ and ${\cal K}$ are, respectively, the superpotential
and the K\"ahler potential of the model, while $\left\{ \phi_i \right\}$
denotes the set of scalar fields amplified during preheating and
rescattering. To quantize the transverse gravitino, we define a
homogeneous mass $m_{3/2}^0$ by replacing in eqn.~(\ref{mgra}) the (${\bf
x}-$dependent) values of the fields with their (homogeneous) variances,
$\phi \left( t , \, {\bf x} \right) \rightarrow \sqrt{ \langle \phi^2
\rangle \left( t \right)}\,$. The difference $\delta m_{3/2} \equiv
m_{3/2}-m_{3/2}^0$ will be accounted for in the interaction lagrangian.
The main production of gravitinos is expected to occur close to the point
at which the number density of light scalars $X$ reaches its maximum, in
the same way as most of the thermal production occurs as soon as the
thermal bath is generated. From the results of section 4, we observe that
the maximum of $N_{c,X}$ is achieved at the end of the rescattering phase.
At this moment the thermalization stage begins, during which the variances
of the fields show an adiabatic evolution. This allows us a consistent
quantization of the gravitino component, since the mass $m_{3/2}^0 \left(
t \right)$ is also varying only adiabatically in this period.
It is clear that the main concern with the procedure just described is
that, contrary to the thermal case, the difference $\delta m_{3/2}$ is now
of the same order of $m_{3/2}^0\,$ itself, at least during the initial
part of the thermalization stage. This leads to the problem discussed in
section 5, namely to the fact that the perturbative computation of the
gravitino production is presumably not as straightforward as in the
thermal case, and more complicated processes than just $2 \rightarrow 2$
interactions can be expected to be relevant. However, as we have already
remarked, the latter should provide at least an order of magnitude
estimate for the gravitino produced from the end of the rescattering stage
on, and should reasonably lead to a lower bound to the total production.
For this reason, we now proceed to a rough estimate of their cross
sections.
The dominant processes with two gravitinos as outgoing particles have two
gravitationally suppressed vertices (i.e. $X \, X \rightarrow \psi_{3/2}
\, \psi_{3/2}$ with a flat direction fermion $x$ in the propagator;
processes coming from the interaction term $\delta m_{3/2} \, {\bar
\psi_{3/2}} \, \psi_{3/2} \,$ are subdominant). Their cross section is of
the order $\sigma \sim p^2 / M_P^4 \,$, where here and in the following
$p$ denotes the typical momentum exchanged in the scattering. As in the
thermal case, the distributions of the light quanta are indeed
characterized by a typical momentum; while for a thermal distribution $p
\sim T\,$, we now have $p \sim \kappa \, \sqrt{\lambda} \, \phi_0\, / a
\left( t \right) \,$, where in the cases shown below $\kappa$ is a
coefficient of order $10$ dependent on the specific choice of the
parameters. In our estimates we will take $\kappa \sim
15\,$.~\footnote{The existence of a typical momentum allows the use of the
integrated Boltzmann equation to estimate the amount of gravitinos
produced. Moreover, since this momentum is much  higher than the gravitino
mass, the value of the latter does not affect the cross sections for the
processes with outgoing transverse gravitinos. This is welcome, since the
above (somewhat arbitrary) redefinition $m_{3/2} = m_{3/2}^0 + \delta
m_{3/2}$ will not affect our estimates.} Thus, $\sigma \sim 10^{-11} \,
M_P^{-2} \, a^{-2}$ for this class of processes.
A more efficient production is expected from scatterings with only one
gravitationally  suppressed vertex, and hence with only one gravitino in
the final state. For example,  the standard thermal production is mainly
due to channels having a gauge interaction  as the second vertex, e.g. $H
\, H \rightarrow \psi_{3/2} \, {\tilde z}$  with an exchanged higgsino. In
the present context, however, processes with outgoing  gauginos (that we
generically denote with ${\tilde g}$)  are expected to be kinematically
forbidden, since these particles acquire a high  effective mass from their
interaction with the nonthermal scalar distributions. Indeed, if a scalar
field $X$ has a large vev, and an interaction of the form $\sqrt{\alpha} X
\psi \tilde{g}$ is present ($\psi$ and $\tilde{g}$ are two component
matter fermion and gaugino, we use $\sqrt{\alpha}$ because we have already
used $g$ as the inflaton-neutrino coupling) then the gaugino acquires a
large Dirac mass $\sim \sqrt{\alpha} \langle X \rangle$ mixing with
$\psi$. We have large variances, rather than a large vev; by analogy with
finite temperature, we expect that $ \langle X^2 \rangle \neq 0$ will
generate an effective mass square ``$m^2 \sim \alpha \langle X^2
\rangle$'' in the $\tilde{g}$ and $\psi$ dispersion relations.\footnote{It
is implicit in this discussion that all the fermionic fields are quantized
in the same way as done for the gravitinos.} So for kinematic purposes, we
assume that gauginos which couple to the flat direction have masses of
order $\sqrt{\alpha  \langle X^2 \rangle}$.
When the light scalar distributions  saturate (that is, when the gravitino
production we are discussing can be effective), we find  numerically
$\sqrt{\langle X^2 \rangle} \sim \left( 10^{-2} - 10^{-1} \right)  \phi_0
/ a \,$. As a consequence, $m_{\tilde g} \sim \left( 10^2 - 10^3 \right)
p\,$,  and these scatterings are forbidden. One is immediately  led to
consider processes with an additional $X_i \, \psi_j \, {\tilde g}$ 
vertex and in which the heavy gaugino is off-shell. Their cross section
can  be roughly estimated as $\sigma \sim 10^{-2} \, \left( \alpha/M_P
\right)^2 \,  \left( p / m_{\tilde g} \right)^2\,$, which is comparable or
smaller than the  cross section for the process $X \, X \rightarrow
\psi_{3/2}  \, \psi_{3/2}$ considered above.
Finally, there is the possibility that the second vertex comes from the
superpotential term (\ref{delw}), also responsible for the Dirac mass term
for the neutrinos. This  can lead to processes of the kind ${\tilde N}_R
\, X \rightarrow x \, \psi_{3/2}\,$ or $X \, X \rightarrow N_R \,
\psi_{3/2}$ ($x$ denoting the fermionic partner of $X\,$; all processes
have in the propagator the fermionic partner of one of the incoming
scalars). The cross sections for these  processes are roughly estimated as
\footnote{In this estimate it is assumed that the  exchanged momentum $p$
is higher or at most comparable with the mass of the r.h. neutrinos. This
is certainly true when most of the gravitinos are produced, i.e. when the
distributions of light quanta $X$ are about to saturate.}  $\sigma \sim
h^2 /M_P^2 \simeq 10^{-13} \, {\tilde h} / M_P^2 \,$.
Thus, unless of a very small coupling $h\,$, the last class of scatterings
has the highest cross section and dominates the production of the
transverse gravitinos. In particular, processes with one incoming $N_R$
quantum are dominant if $N_{c,X}$ starts to saturate at a smaller value
than $N_{c,N}\,$. Viceversa, scatterings of the kind $ X \, X \rightarrow
N_R \, \psi_{3/2}$ will dominate. Numerical results indicate that the
former situation is more often realized. The cases in which the opposite
was found are characterized by a relatively high coupling ${\tilde h}\,$,
so that the light degrees of freedom are quickly amplified to values
$N_{c,X} \ga N_{c,N}\,$. Anyhow, in these cases the cross sections of the
two type of processes are clearly of the same order of magnitude. Hence,
for brevity of exposition we will only refer to the processes with an
incoming r.h. sneutrino, although both the two possibilities have been
considered in our estimates.
The integrated Boltzmann equation reads
\begin{equation}
\frac{d N_{3/2}}{d t} + 3 \, H \, N_{3/2} \simeq \langle \sigma \vert v \vert \rangle N_{X} N_{N}\,  \,\,,
\label{boltz}
\end{equation}
where the ``friction term'' due to the expansion of the Universe can be
neglected in the estimate of the order of magnitude of gravitinos
produced. The whole production time can be then divided in a series of
time intervals of duration $H^{-\,1} \left( t_i \right)$ each. During each
interval, quanta of gravitinos are generated with a density of
\begin{equation}
\delta N_{3/2}^i \sim \frac{h^2}{M_P^2} \, \frac{N_{c,X} \left( \eta_i \right)}{a \left( \eta_i \right)^3} \, \frac{N_{c,N} \left( \eta_i \right)}{a \left( \eta_i \right)^3} \, H^{-1} \left( \eta_i \right)
\label{granum1}
\end{equation}
(notice the presence of the scale factor, since the physical and not the
comoving occupation number has to be used in the integrated Boltzmann
equation). The function $\left( N_{c,X} \, N_{c,N} \, H^{-1} \, a^{-6}
\right) \left( \eta \right)$ amounts to zero at the end of inflation, and
it reaches a maximum at a time $\eta_*\,$, which can be determined
numerically and which roughly corresponds to the moment at which the
comoving number density $N_{c,X}$ starts saturating (this in turns occurs
towards the end of the rescattering stage, when $2 \rightarrow 2$
processes start dominating). At $\eta > \eta_*$ it then quickly decreases
due to the expansion of the Universe. It thus turns out that, as for the
thermal production, the gravitino quanta are mostly generated at the time
$\eta_*\,$, so that their ``late time'' physical number density is
approximatively given by
\begin{widetext}
\begin{equation}
N_{3/2} \left( \eta \right) \sim \frac{h^2}{M_P^2} \frac{N_{c,X} \left( \eta_* \right)}{a \left( \eta_* \right)^3} \frac{N_{c,N} \left( \eta_* \right)}{a \left( \eta_* \right)^3} H^{-1} \left( \eta_* \right) \left[ \frac{a \left( \eta_* \right)}{a \left( \eta \right)} \right]^3, \;\;\;\;\; \eta > \eta_* \,.
\label{granum2}
\end{equation}
\end{widetext}
This is eqn.~(\ref{granum3}) of the main text.
\end{document}